\documentstyle[12pt]{article}

\input{pstricks}

\renewcommand{\theequation}{\arabic{section}.\arabic{equation}}

\newcounter{saveeqn}
\newcommand{\alpheqn}{\setcounter{saveeqn}{\value{equation}}%
\stepcounter{saveeqn}
\setcounter{equation}{0}%
\renewcommand{\theequation}{\mbox{\arabic{section}.\arabic{saveeqn}
\alph{equation}}}}
\newcommand{\reseteqn}{\setcounter{equation}{\value{saveeqn}}%
\renewcommand{\theequation}{\arabic{section}.\arabic{equation}}}
\newtheorem{proposition}{Proposition}[section]

\newtheorem{theorem}[proposition]{Theorem}
\newtheorem{corollary}[proposition]{Corollary}

\newtheorem{proofs}{Proof}

\newenvironment{proof}{\begin{sloppypar}\begin{proofs}\rm}{\hspace*{\fill}
$\Box$ \end{proofs} \end{sloppypar}}
\newtheorem{lempro}{Proof of the Lemma}

\newtheorem{remar}{Remark}

\newenvironment{remark}{\begin{remar} \rm}{\end{remar}}
\newtheorem{remars}{Remarks}

\newenvironment{remarks}{\begin{remars}\hfill
\rm\begin{enumerate}}{\end{enumerate} \end{remars}}
\newtheorem{conse}{Consequences}

\newtheorem{exams}{Examples}

\newfont{\BigBbb}{msbm10 at 12pt}
\newcommand{\bigBbb}[1]{\mbox{\BigBbb #1}}
\newfont{\SBbb}{msbm8}

\newfont{\euft}{eufm10 at 12pt}
\newcommand{\eufm}[1]{\mbox{\euft #1}}

\newcommand{\diff}{{\rm d}}

\newcommand{\C}{\bigBbb{C}}

\newcommand{\R}{\bigBbb{R}}
\newcommand{\Z}{\bigBbb{Z}}
\newcommand{\A}{\eufm{A}}

\newcommand{\varPi}{{\mit\Pi}}

\begin{document}

\begin{titlepage}

\hfill LMU-TPW 98-10
\vspace{5ex}
\begin{center}
{\Large
{\bf Physically Realistic Solutions to the Ernst Equation\\
on Hyperelliptic Riemann Surfaces}\\[4ex]
\large C.~Klein\\
{\em Institut f\"ur Theoretische Physik, Universit\"at T\"ubingen,\\
Auf der Morgenstelle 14, 72076 T\"ubingen, Germany}\\[2ex]
O.~Richter\\
{\em Sektion Physik der Universit\"at M\"unchen,\\
Theresienstra{\ss}e 37, 80333 M\"unchen, Germany}\\[4ex]
June 10, 1998}%
\end{center}
\vspace{3ex}
\begin{abstract}
We show that the class of hyperelliptic solutions to the Ernst equation 
(the stationary axisymmetric 
Einstein equations in vacuum) previously discovered by 
Korotkin and Neugebauer and Meinel
can be  derived via Riemann--Hilbert 
techniques. The present paper extends the discussion of the physical 
properties of these 
solutions that was begun in a Physical Review Letter, and supplies complete 
proofs. We identify a physically 
interesting subclass where the Ernst potential is everywhere 
regular except at  a closed surface which might be identified with the 
surface of a body of revolution. The corresponding spacetimes are 
asymptotically flat and equatorially symmetric. This suggests that they could 
describe the 
exterior of an isolated body, for instance a relativistic star or a 
galaxy. Within this class, one has 
the freedom to specify a real function and a set of complex parameters 
which can possibly be used to solve certain boundary value problems for the 
Ernst equation. The solutions can have ergoregions, a Minkowskian limit and 
an ultrarelativistic 
limit where the metric approaches the extreme Kerr solution. We give 
explicit formulae for the potential on the axis and in the equatorial 
plane where the expressions simplify. 
Special attention is paid to the simplest non--static solutions (which are of 
genus two) to which the rigidly rotating dust disk belongs.
\end{abstract}
\vspace{3ex}
\centerline{PACS numbers: 04.20.Jb, 02.10.Rn, 02.30.Jr}
\end{titlepage}

\section{Introduction}\label{sec1}

It is generally believed that most of the stars and galaxies can be described 
in good approximation as fluid bodies in thermodynamical equilibrium. In 
the framework of general relativity, this implies (see e.g.\ 
\cite{hartle,lindblom}) that the corresponding spacetimes are stationary and 
axisymmetric. Moreover it is usually assumed (though there is no proof 
known to us) that they are equatorially symmetric.  
This stresses the importance of the study of stationary axisymmetric 
spacetimes. A relativistic treatment is necessary for rapidly 
rotating and massive compact objects like pulsars, neutron stars and 
black-holes. 

Though the importance of global solutions describing stationary 
axisymmetric fluid bodies is generally accepted, the complicated structure 
of the Einstein equations with matter gives little hope that such solutions 
can be found in the near future. Only for special and somewhat unphysical 
equations of state 
\cite{wahl,kramer,seno}, it was possible to give solutions in the 
matter region which are discussed as candidates for an interior solution. 
In the exterior vacuum region, however, powerful solution generating techniques 
are at hand. Since the surface $\Gamma_z$
of a compact astrophysical object constitutes a 
natural boundary at which the metric functions are not continuously 
differentiable, one is looking 
for solutions to the vacuum equations
that are analytic outside this contour and can be at least 
continuously extended to $\Gamma_z$. This means that the typical problem 
one has to consider for the vacuum Einstein equations is a boundary value 
problem of Dirichlet, von Neumann or mixed type, see \cite{hp2}.  The 
matter then enters only in form of boundary conditions for the vacuum 
equations.  This is possible if an interior solution is known or if only 
two--dimensionally extended bodies like disks or shells are considered. 
In the latter case,
the surfacelike distribution of the matter implies 
that the matter equations reduce to ordinary differential equations. 
Notice that disks are important models in astrophysics for certain types 
of galaxies.

The reason why it is much more promising to treat only the vacuum case 
is the equivalence of the stationary axisymmetric Einstein equations 
to a single nonlinear differential equation for a complex 
potential, the so called 
Ernst equation \cite{ernst}. The latter belongs to a family of 
completely integrable nonlinear equations that are studied as
the integrability conditions for associated linear differential systems. The 
common feature of these linear systems is that they contain an additional 
variable, the so called spectral parameter, which reflects an underlying 
symmetry of the differential equations under investigation, in the case of 
the Ernst equation the Geroch group \cite{geroch}. 
Associated linear systems for the Ernst equations 
were given in \cite{belzak,maison,neuglinear}. 
The existence of this parameter can be used to construct solutions by
prescribing the singular structure of the matrix of the linear system with
respect to the spectral parameter.

One of the most successful solution techniques for nonlinear differential
equations rests on methods of
algebraic geometry and leads to the so called finite-gap solutions that can be
expressed elegantly in terms of theta functions. Such methods were first used
to construct periodic and quasiperiodic
solutions to nonlinear evolution equations like the Korteweg--de~Vries 
(KdV) and the Sine--Gordon (SG) equation.  For a
survey of this subject we refer the reader to~\cite{soliton1,algebro}. 
However, it was only recently that 
algebro-geometrical methods were applied to the Ernst equation,
see~\cite{korot1}. The found solutions differ from similar 
solutions of other equations in several aspects, e.g.\ they are in general 
not  periodic or quasi-periodic. The main difference is that 
this class is much richer than previously obtained ones.  

The development of solution techniques yields a deeper insight into the
structure of nonlinear differential equations. However, from a practical point
of view, it would also be desirable to solve initial value problems or, for the
Ernst equation, boundary value problems.
One approach to solve boundary value problems of the above mentioned
type with the help of the linear system is to translate the 
physical boundary conditions into a Riemann-Hilbert problem which is
equivalent to a linear integral equation, see~\cite{soliton1}.
Neugebauer and Meinel~\cite{ngbml1}
succeeded in doing this in the case of the rigidly rotating dust disk.  They
were able to reduce the matrix problem on a sphere to a scalar
Riemann-Hilbert problem
on a hyperelliptic Riemann surface which can be solved explicitly via 
quadratures. By making use of the gauge transformations of the linear system
we were able to show~\cite{prd} that this is possible in
general if the boundary value problem leads to a Riemann-Hilbert
problem with rational jump data. Up to now there is however no direct 
way to infer the jump data from the boundary value problem one wants to solve.
The explicit form of the hyperelliptic solutions possibly offers a different
approach to boundary value problems: one can try to identify the free 
parameters in the solutions, a real valued function and a set of complex
parameters, the branch points of the hyperelliptic Riemann surface, from 
the problem one wants to solve. 

To this end we study a class of solutions -- which is essentially
equivalent to \cite{korot1} and \cite{meinelneugebauer} --
that is constructed via a  
generalized Riemann--Hilbert problem on a hyperelliptic Riemann surface. 
We present a complete  discussion of the singularity
structure  of these Ernst potentials. It is possible to 
identify a subclass of solutions that are everywhere regular except  
at some contour, which can possibly be related to the surface of an
isolated body, where the Ernst potential is bounded. These solutions 
are asymptotically flat and
equatorially symmetric, and thus show all the features one might expect 
from the exterior solution for an isolated relativistic ideal fluid. 
They can have a Minkowskian and an extreme relativistic limit in 
which the body is `hidden' behind a horizon, and the exterior solution 
becomes the extreme Kerr solution. This provides the hope that further 
solutions to physically interesting boundary value problems to the Ernst
equation, besides the rigidly rotating dust disk, can be identified within this
class. First results on this subclass where published in \cite{prl}.

The paper is organized as follows. In section~\ref{sec2} we introduce the 
linear system associated to the Ernst equation and  discuss how the matrix 
of the system has to be constructed in order to end up with new solutions 
to the Ernst equation. Using the results of~\cite{krichever1}, we show how
Riemann surfaces arise naturally in the context of linear systems with a
spectral parameter. In the case of the Ernst equation, these are hyperelliptic
Riemann surfaces with a special structure of the branch points. We will
restrict ourselves to regular compact Riemann surfaces and are eventually led
to consider families of hyperelliptic Riemann surfaces of arbitrary genus,
parametrized by the physical coordinates.

In section 3 we recall some basic notions of the theory of Riemann surfaces,
theta functions and the solution of Riemann--Hilbert problems on
Riemann surfaces due to Zverovich, and present the class of solutions. It 
is shown that the solution of the axisymmetric Laplace equation which can 
be freely prescribed in \cite{meinelneugebauer}
is a period of the Abelian integrals 
which determine the singularity structure of the matrix of the linear 
system. The differential relations between these periods are a subset of 
the so called Picard--Fuchs equations which we write down for the Ernst 
equation.
In section 4 we discuss the singularity structure of these
solutions. It is shown that the solutions can have a regular axis and 
are in general asymptotically flat. Using an identity for 
theta functions, we are able to give in section 5 
compact formulas for two metric 
functions and a simple condition for the occurrence of ergospheres.
A subclass of solutions with equatorial symmetry is presented in section 6. 
The common physical features of this subclass like  the 
extreme relativistic limit are discussed.  
In section 7, we use the equatorial symmetry to give simplified formulae 
for the potential in the equatorial plane  and on the axis. Since the rigidly 
rotating dust disk belongs to the simplest non-static solutions which are
of genus 2, we consider this case in detail in section 8. In section 9, we 
summarize the results and add some concluding remarks.

\section{Linear System for the Ernst equation and 
Monodromy matrix}\label{sec2}
\setcounter{equation}{0}

It is well known (see \cite{exac}) that the metric of stationary axisymmetric 
vacuum spacetimes can be written in the  Weyl--Lewis--Papapetrou form
\begin{equation}\label{3.1}
	\diff s^2 =-{\rm e}^{2U}(\diff t+a\diff \phi)^2+{\rm e}^{-2U}
	\left({\rm e}^{2k}(\diff \rho^2+\diff \zeta^2)+
	\rho^2\diff \phi^2\right)
	\label{vac1}
\end{equation}
where $\rho$ and $\zeta$ are Weyl's canonical coordinates and 
$\partial_{t}$ and $\partial_{\phi}$ are the two commuting asymptotically
timelike respectively spacelike Killing vectors. 

In this case the vacuum field equations are equivalent 
to the Ernst equation for the 
complex potential $f$ where $f={\rm e}^{2U}+{\rm i}b$, and where
the real function $b$ 
is related to the metric functions via
\begin{equation}\label{3.2}
	b_{,z}=-\frac{{\rm i}}{\rho}{\rm e}^{4U}a_{,z}
	\label{vac9}.
\end{equation}
Here the complex variable $z$ stands for $z=\rho+{\rm i}\zeta$. With these
settings, the Ernst equation reads
\begin{equation}\label{3.3}
	f_{z\bar{z}}+\frac{1}{2(z+\bar{z})}(f_{\bar{z}}+f_z)=\frac{2 }{f+\bar{f}}
	f_z f_{\bar{z}}
        \label{vac10}\enspace,
\end{equation}
where a bar denotes complex conjugation in $\C$. With a solution $f$,
the metric function $U$ follows directly from the definition of the Ernst 
potential whereas $a$ can be obtained from (\ref{vac9}) via quadratures. 
The metric function $k$ can be calculated from the relation
\begin{equation}\label{3.4}
		k_{,z}  =  2\rho \left(U_{,z}\right)^2-\frac{1}{2\rho}{\rm e}^{4U}
                \left(a_{,z}\right)^2.
	\label{vac8}
\end{equation}
The integrability condition of (\ref{vac9}) and  (\ref{vac8}) is the 
Ernst equation.

The remarkable feature of the Ernst equation is that it is completely 
integrable. This means that it can be considered as the integrability 
condition of an overdetermined linear differential system for a 
matrix valued function $\Phi$ that contains an additional variable, the so 
called spectral parameter $K$. The occurrence of the linear system with a 
spectral parameter is a consequence of the symmetry group of the Ernst 
equation, the Geroch group \cite{geroch}. Several forms of the linear system 
are known 
in the literature (\cite{belzak,maison,neuglinear}). They are related 
through gauge transformations (see \cite{cosgrove}). The choice of a 
specific form of the linear system is equivalent to a gauge fixing. We 
will use the form of \cite{neuglinear},
\alpheqn
\begin{eqnarray}
	\Phi_{,z}(K,\mu_0;z,\bar{z})& = & \left\{\left(
	\begin{array}{cc}
		N & 0  \\
		0 & M
	\end{array}
	\right)
        +\frac{K-{\rm i}\bar{z}}{\mu_0(K)}\left(
	\begin{array}{cc}
		0 & N  \\
		M & 0
	\end{array}
        \right)\right\}\Phi(K,\mu_0;z,\bar{z})\doteq W\Phi\enspace,
	\label{lin1} \\
	\Phi_{,\bar{z}}(K,\mu_0;z,\bar{z}) & = & \left\{\left(
	\begin{array}{cc}
		\bar{M} & 0  \\
		0 & \bar{N}
	\end{array}
	\right)
        +\frac{K+{\rm i}z}{\mu_0(K)}\left(
	\begin{array}{cc}
		0 & \bar{M}  \\
		\bar{N} & 0
	\end{array}
        \right)\right\}\Phi(K,\mu_0;z,\bar{z})\doteq V\Phi
	\label{lin2}
\end{eqnarray}
\reseteqn
where
\begin{equation}\label{3.6}
	M  =  \frac{f_z}{f+\bar{f}}, \quad
	N  =  \frac{\bar{f}_z}{f+\bar{f}}.
\end{equation}
Obviously $M$ and $N$ depend only on the coordinates $z$ and $\bar{z}$ 
and not on the spectral parameter $K$ that lives on the Riemann surface
${\cal L}(z,\bar{z})={\cal L}$  given by $\mu_0^2(K)=(K-{\rm i}
\bar{z})(K+{\rm i}z)$. Notice that
${\cal L}$ is a Riemann surface of genus zero with coordinate dependent 
branch points. This is a special feature of the family of chiral field 
equations to which the Ernst equation belongs that has no counterpart among 
the completely integrable nonlinear evolution equations for which 
algebro-geometric solutions have been constructed first.

On $\cal L$ we have an involutive map $\sigma$, defined by
\begin{equation}\label{3.61}
{\cal L}\ni P=(K,\pm\sqrt{(K-{\rm i}\bar{z})(K+{\rm i}z)})\to\sigma(P)\equiv
P^{\sigma}=(K,\mp\sqrt{(K-{\rm i}\bar{z})(K+{\rm i}z)})\in{\cal L}\enspace,
\end{equation}
and an anti--holomorphic involution $\tau$, defined by
\begin{equation}\label{3.62}
{\cal L}\ni P=(K,\pm\sqrt{(K-{\rm i}\bar{z})(K+{\rm i}z)})\to\tau(P)\equiv
\bar{P}=(\bar{K},\pm\sqrt{(\bar{K}-{\rm i}\bar{z})(\bar{K}+
{\rm i}z)})\in{\cal L}\enspace.
\end{equation}

It is possible to use the existence of the above linear system for the 
construction of solutions to the Ernst equation. To this end one 
investigates the singularity structure of the matrices $\Phi_{z}\Phi^{-1}$ and 
$\Phi_{\bar{z}}\Phi^{-1}$ with respect to the spectral parameter and infers 
a set of conditions for the  matrix $\Phi$ 
(at least twice differentiable with respect 
to $z$, $\bar{z}$) that satisfies the linear system 
(\ref{lin1}) and (\ref{lin2}). 
This is done (see e.g.~\cite{korot1}) in
\begin{theorem}\label{theorem2.1}
 Let $\Phi(P)$ ($P\in{\cal L}$) be a $2\times 2$--matrix with the following
properties:\\
I. $\Phi(P)$  is holomorphic and invertible at the branch
points $P_0=-{\rm i}z$ and $\bar{P}_0$ such that the logarithmic derivative
$\Phi_{z}\Phi^{-1}$ diverges as $(K+{\rm i}z)^{\frac{1}{2}}$ at $P_0$ and 
$\Phi_{\bar{z}}\Phi^{-1}$ as $(K-{\rm i}\bar{z})^{\frac{1}{2}}$ at
$\bar{P}_0$.\\
II. All singularities of $\Phi$ on ${\cal L}$ (poles, essential
singularities, zeros of the determinant of $\Phi$, branch cuts and branch
	points) are regular which means that the logarithmic derivatives 
	$\Phi_{z}\Phi^{-1}$ and $\Phi_{\bar{z}}\Phi^{-1}$ are holomorphic in the 
        neigbourhood of the singular points (this implies they have to be
        independent
        of $z$, $\bar{z}$). In particular $\Phi(P)$ should have\\
	a) regular singularities at the points $A_i\in {\cal L}$ ($i=1,\dots,n$)
        which do not depend on $z$, $\bar{z}$,\\
	b)  regular essential singularities at the points $S_i$ 
        ($i=1,\dots,m$) which do not depend on $z$, $\bar{z}$,\\
        c)  boundary values at a set of (orientable, piecewise smooth)
	contours $\Gamma_i \subset {\cal L}$ ($i=1,\dots,l$)  
        independent of $z$, $\bar{z}$, which are related on both sides of the
        contours via
	\begin{equation}\label{3.9}
\left.\Phi_- (P)=\Phi_+ (P) {\cal G}_i(P)\right|_{P\in\Gamma_i}\enspace.
          \label{lin6}
	\end{equation}
	where  ${\cal G}_i(P)$ are matrices independent of $z$, $\bar{z}$ with 
        H\"older--continuous components and non--vanishing determinant.\\
III. $\Phi$ satisfies the reduction condition
	\begin{equation}\label{3.10}
	        \Phi(P^{\sigma}) = \sigma_3 \Phi(P) \gamma\enspace,
		\label{lin7}
	\end{equation}
	where $\sigma_3$ is the third Pauli matrix, and where $\gamma$ is an 
        invertible matrix independent of $z$ and $\bar{z}$.\\
	IV. The normalization and reality condition
\begin{equation}
	\Phi(P=\infty^+)=\left(
	\begin{array}{rr}
		\bar{f} & 1  \\
		f & -1
	\end{array}
	\right)
	\label{lin9}.
\end{equation}

Then the function $f$ in (\ref{lin9}) is a solution to the Ernst equation.
\end{theorem}
A proof of this Theorem may be obtained by comparing the above matrix
$\Phi$ with the linear system (\ref{lin1}) and (\ref{lin2}).
\begin{proof}
Because of I, $\Phi$ and $\Phi^{-1}$ can be expanded in a series in
$t=\sqrt{K+iz}$ and $t'=\sqrt{K-i\bar{z}}$ in a neighbourhood of $P=P_0$
and $P=\bar{P}_0\neq P_0$ respectively at all points $P_0$, $\bar{P}_0$ which 
do not belong to the singularities given in II. This implies that $\Phi_z
\Phi^{-1}=\alpha_0/t+\alpha_1+\alpha_2t +\cdots$. We recognize that, because of
I and II, $\Phi_z \Phi^{-1}-\alpha_0/t$ is a holomorphic function. The
normalization condition IV implies that this quantity is bounded at 
infinity. According to Liouvilles theorem, it is a constant. 
Since $\Phi$, $\Phi^{-1}$ and $\Phi_z$ are single valued functions on ${\cal 
L}$, they must be functions of $K$ and $\mu_0$. Therefore we have $\Phi_z
\Phi^{-1}=\beta_0 \sqrt{\frac{K-\bar{P}_0}{K-P_0}}+\beta_1$. The matrix 
$\beta_0$ must be 
independent of $K$ and $\mu$ since $\Phi_z \Phi^{-1}$ must have the same 
number of zeros and poles on ${\cal L}$. The structure of the matrices 
$\beta_0$ and $\beta_1$ follows from III. From the normalization condition IV,
it follows that $\Phi_z \Phi^{-1}$ has the structure of (\ref{lin1}).
The corresponding equation for $\Phi_{\bar{z}}\Phi^{-1}$ can be obtained 
in the same way.
\end{proof}

For a given Ernst potential $f$, the matrix $\Phi$ in the above theorem is 
not uniquely determined. This reflects the fact that the gauge is not uniquely 
fixed in the linear system (\ref{lin1}) and (\ref{lin2}). 
If we choose without loss of generality 
$\gamma=\sigma_1$ (the first Pauli matrix), 
the remaining gauge freedom can be seen from 
\begin{corollary}
        Let $\Phi(P)$ be a matrix subject to the conditions of
Theorem~\ref{theorem2.1},
and $C(K)$ be a $2\times2$-matrix that only depends on $K\in \C$
with the properties
\begin{eqnarray}
	C(K) & = & \alpha_1(K) \hat{1}+\alpha_2(K) \sigma_1,
	\nonumber \\
	\alpha_1(\infty) &=  &1,\quad \alpha_2(\infty)=0.
	\label{c}
\end{eqnarray}
Then the matrix $\Phi'(P)=\Phi(P) C(K)$ also satisfies the conditions of
Theorem~\ref{theorem2.1} and $\Phi'(\infty^+)=\Phi(\infty^+)$.
\end{corollary}
It is this gauge freedom to which we refer when we speak of the gauge 
freedom of the linear system in the following.

It is interesting to note that the metric function $a$ can be obtained 
from a given matrix $\Phi$ without solving the equation (\ref{vac9}), see 
\cite{korot1}. We get  
\begin{proposition}
        Let $\delta$ be a local parameter in the vicinity of $\infty^-$. Then
	\begin{equation}
                (a-a_0)e^{2U}={\rm i}(\Phi_{11}-\Phi_{12})_{,\delta}\enspace,
		\label{a}
	\end{equation}
	where $a_0$ is a constant that is fixed by the condition that $a=0$ on 
	the regular part of the axis and at spatial infinity, and where 
	$\Phi_{,\delta}$ denotes the linear term in the expansion of $\Phi$ in 
	$\delta$ divided by $\delta$. 
\end{proposition}
The proof follows from the linear system (\ref{lin1}) and (\ref{lin2}).
\begin{proof}
	It is straightforward to check the relation 
	\begin{equation}
		(\Phi^{-1}\Phi_{,\delta})_{,z}=\Phi^{-1}(\Phi_{z}\Phi^{-1})_{,\delta} 
\Phi
                \label{a1}\enspace.
	\end{equation}
With  (\ref{lin1}), we get 
\begin{equation}
\left(\Phi^{-1}\Phi_{\delta}\right)_{21,z}=\frac{{\rm i}\rho}{(f+\bar{f})^2}
        (\bar{f}-f)_z\enspace,
	\label{a2}
\end{equation}
from which, together with (\ref{vac9}), (\ref{a}) follows.
\end{proof}
Notice that $a_0$ is not gauge independent (in the sense of the above 
corollary) whereas $a$ is.

Theorem~\ref{theorem2.1} can be used to construct solutions to the
Ernst equation by determining the structure and the singularities of 
$\Phi$ in accordance with the conditions I--IV. For nonlinear evolution
equations, large classes of solutions were constructed with the help of
algebro-geometric methods, in particular Riemann surface techniques. A 
keypoint in this context is the occurrence of Riemann surfaces
which are related to the linear system of the integrable equation under
consideration. In this paper we want to show how solutions for the Ernst 
equation
can be constructed by making use of the so called monodromy matrix
of the Ernst system, which -- following~\cite{krichever1} --  can be introduced
as follows.

For a given linear system (\ref{lin1}) and (\ref{lin2}), we define the
monodromy matrix $L$ as a solution to the system
\begin{equation}\label{3.19}
        L_z=[W,L],\quad L_{\bar{z}}=[V,L]
        \label{mono3}\enspace.
\end{equation}
For a known solution $\Phi$ of (\ref{lin1}) and (\ref{lin2}), 
$L$ can be directly constructed in the form
\begin{equation}\label{3.20}
        L(K)=-\hat{\mu}(K) \Phi{\cal C}\Phi^{-1}
        \label{mono5}
\end{equation}
where ${\cal C}$ is an arbitrary constant matrix with $\det{\cal C}=-1$ and
$\hat{\mu}$ does not depend on the physical coordinates. Since $\Phi$
is analytic in $K$, there is a solution to (\ref{mono3}) with the same 
properties.

It follows from (\ref{mono3}) that the coefficients of the characteristic 
polynomial $Q(\mu, K)=\det (L(K)-\hat{\mu}\hat{1})$ are independent of the
coordinates. Without loss of generality we may assume $\mbox{Tr}L(K)=0$. Then
$L$ has the structure
\begin{equation}\label{3.21}
        L=\left(
        \begin{array}{rr}
                A(K) & B(K)  \\
                C(K) & -A(K)
        \end{array}
        \right)
        \label{mono4}\enspace.
\end{equation}
The equation $Q(\hat{\mu},K)=0$, i.e. 
\begin{equation}\label{3.21.1}
        \hat{\mu}^2=A^2+BC\enspace,
        \label{mono6}
\end{equation}
is then the equation of an algebraic curve which 
in general will have infinite genus. We will restrict the analysis in the 
following to the case of a regular curve with finite genus.

In this case, the Riemann surface $\hat{{\cal L}}$ is given by an 
equation of the form  
\begin{equation}\label{3.22}
        \hat{\mu}^2=\prod_{i=1}^{g}(K-E_i)(K-F_i)
        \label{mono20}
\end{equation}
where $E_i$ and $F_i$
are obviously independent of the physical coordinates. This equation represents
a two sheeted covering of the Riemann sphere and thus a four sheeted covering
of the complex plane. A point $\hat{P}\in\hat{\cal L}$ can be given by
$\hat{P}=(K,\mu_0(K),\hat{\mu}(K))$. The Hurwitz diagram of $\hat{{\cal L}}$ 
is shown in figure 1.
\begin{figure}[ht]
\begin{center}
\unitlength1cm
\begin{picture}(7,4)
\thicklines
\put(0.4,4){\line(1,0){6.6}}
\put(0.4,3){\line(1,0){6.6}}
\put(0.4,2){\line(1,0){6.6}}
\put(0.4,1){\line(1,0){6.6}}
\put(0.9,4){\line(0,-1){2}}
\put(1.2,3){\line(0,-1){2}}
\put(2.2,4){\line(0,-1){2}}
\put(2.5,3){\line(0,-1){2}}
\multiput(3.7,4)(0.8,0){2}{\line(0,-1){1}}
\multiput(3.7,2)(0.8,0){2}{\line(0,-1){1}}
\multiput(5.7,4)(0.8,0){2}{\line(0,-1){1}}
\multiput(5.7,2)(0.8,0){2}{\line(0,-1){1}}
\put(0.9,4){\circle*{0.15}}
\put(0.9,2){\circle*{0.15}}
\put(1.2,3){\circle*{0.15}}
\put(1.2,1){\circle*{0.15}}
\put(2.2,4){\circle*{0.15}}
\put(2.2,2){\circle*{0.15}}
\put(2.5,3){\circle*{0.15}}
\put(2.5,1){\circle*{0.15}}
\multiput(3.7,4)(0.8,0){2}{\circle*{0.15}}
\multiput(3.7,3)(0.8,0){2}{\circle*{0.15}}
\multiput(3.7,2)(0.8,0){2}{\circle*{0.15}}
\multiput(3.7,1)(0.8,0){2}{\circle*{0.15}}
\multiput(5.7,4)(0.8,0){2}{\circle*{0.15}}
\multiput(5.7,3)(0.8,0){2}{\circle*{0.15}}
\multiput(5.7,2)(0.8,0){2}{\circle*{0.15}}
\multiput(5.7,1)(0.8,0){2}{\circle*{0.15}}
\put(0,3.9){4}
\put(0,2.9){3}
\put(0,1.9){2}
\put(0,0.9){1}
\put(0.9,0.5){${\rm i}\bar{z}$}
\put(2,0.5){$-{\rm i}z$}
\put(3.5,0.5){$E_1$}
\put(4.3,0.5){$F_1$}
\put(4.9,0.5){$\cdots$}
\put(5.5,0.5){$E_g$}
\put(6.3,0.5){$F_g$}
\put(3.5,2.5){$E_1^\sigma$}
\put(4.3,2.5){$F_1^\sigma$}
\put(4.9,2.5){$\cdots$}
\put(5.5,2.5){$E_g^\sigma$}
\put(6.3,2.5){$F_g^\sigma$}
\end{picture}
\end{center}
\caption{The Hurwitz diagram of $\hat{\cal L}$.}
\end{figure}
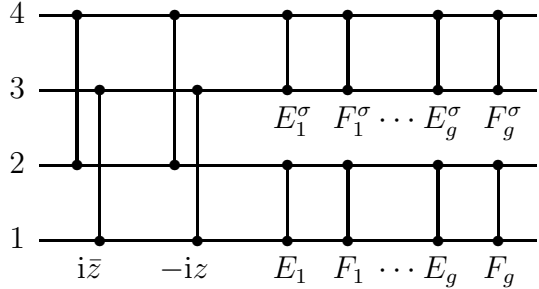

There is an automorphism $\sigma$ of $\hat{\cal L}$ inherited from ${\cal L}$
which ensures $E_i^{\sigma}=E_i$ and $F_i^{\sigma}=F_i$. The orbit space
${\cal L}_H=\hat{\cal L}/\sigma$ is then, see~\cite{algebro}, again a Riemann
surface, namely a hyperelliptic surface given by
\begin{equation}\label{3.22a}
\mu_H^2=(K-{\rm i}\bar{z})(K+{\rm i}z)\prod_{i=1}^{g}(K-E_i)(K-F_i)\enspace.
\end{equation}
Thus it is possible to construct components of the matrix $\Phi$ on ${\cal 
L}_H$ which makes it possible to use the powerful calculus of hyperelliptic 
Riemann surfaces. These functions may be lifted to $\hat{{\cal L}}$. 
As we will show in the following, it is possible to construct a 
matrix $\Phi$ on ${\cal L}$ in accordance with the conditions of Theorem 2.1 
by projecting onto this surface.

\section{Hyperelliptic solutions of the Ernst equation}
\setcounter{equation}{0}

\subsection{Theta functions asscociated with a Riemann surface and the
Riemann--Hilbert problem}

In this section, we want to give an explicit construction of the matrix 
$\Phi$ in accordance with Theorem 2.1. Condition II can be used to 
construct solutions by 
prescribing the poles, essential singularities and cuts of $\Phi$ 
which is equivalent to the solution of 
a generalized Riemann--Hilbert problem for the matrix $\Phi$. The 
investigation of such matrix Riemann--Hilbert problem  turns out to be 
rather difficult and is not yet fully done (in general it can be merely 
reduced to the solution of a linear integral equation, see e.g.\ 
\cite{musk}). Therefore we will use here a different approach. The 
occurrence of the monodromy matrix suggests that it might be possible to 
construct a matrix $\Phi$ on the Riemann surface $\hat{{\cal L}}$ of the 
previous section. The additional freedom we thus gain is used to restrict 
the problem to a scalar one, namely to a Riemann--Hilbert problem for one 
component of $\Phi$ on the hyperelliptic surface ${\cal L}_H$ obtained 
from $\hat{{\cal L}}$ by factorizing with respect to the involution 
$\sigma$.  We impose the reality condition $E_i, F_i\in \R$ or 
$E_i=\bar{F}_i$ on the branch points in order to satisfy the reality 
condition of Theorem 2.1. 
Then we construct the whole matrix $\Phi$ in accordance with 
this theorem. In fact it was shown in \cite{prd} that all matrix 
Riemann--Hilbert problems with rational jump data are gauge equivalent 
to scalar problems on a suitably chosen hyperelliptic surface. Thus the 
limitation to the scalar case is only a comparatively weak restriction 
which allows, as we will show below, for an explicit solution of the 
problem in terms of theta functions. 

For the moment, we fix the physical coordinates $z$ and 
$\bar{z}$ in a way that $\rho\neq0$ and that $-{\rm i}z$ and ${\rm i}\bar{z}$
do not coincide with the singular points of 
$\Phi$ in order to ensure that the first condition of Theorem 2.1 is valid.
In the next section we study the dependence of the found solution on
$z$ and $\bar{z}$.
In order to give the solution to this special case of the generalized
Riemann--Hilbert problem, we use the theory of theta functions associated to
a Riemann surface (see \cite{dubrovin}) and the solution of the Riemann--Hilbert problem on a Riemann
surface, as given in~\cite{zverovich1}.
As we will need only hyperelliptic Riemann surfaces
of the form (\ref{3.22a}), we restrict ourselves to this case.

Let us denote by $(a_1,\dots,a_g,b_1,\dots,b_g)$ a basis of the
first (integral) homology group $H_1({{\cal L}_H})$ of ${{\cal L}_H}$ (see 
the picture below) where  the
cuts are either between real branch points (which are ordered 
$E_{k+1}<F_{k+1}<\ldots $) or between $E_i$ and $\bar{E}_i$ (for the moment 
we ignore the case that more than two branch points may have the same real 
part).\\
\begin{figure}[ht]
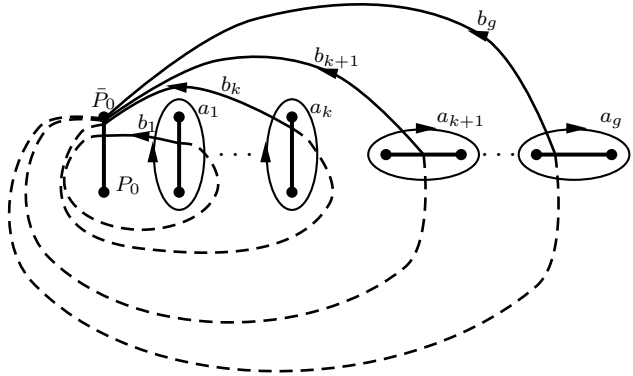

\setlength{\unitlength}{.5cm}
\begin{center}
\psset{unit=.5cm}
\pspicture[](-2,0)(14,6)
\psline[linewidth=1.5pt]{*-*}(0,-1)(0,1)
\rput[bl](.3,-1.05){$\scriptstyle P_0$}
\rput[b](0,1.2){$\scriptstyle\bar{P}_0$}
\psellipse[](2,0)(.7,1.5)
\psdots*[dotstyle=triangle*,dotscale=1.2 3](1.3,0)
\rput[b](2.8,1){$\scriptstyle a_1$}
\psline[linewidth=1.5pt]{*-*}(2,-1)(2,1)
\rput(3.5,0){$\dots$}
\psellipse[](5,0)(.7,1.5)
\psdots*[dotstyle=triangle*,dotscale=1.2 3](4.3,0)
\rput[b](5.8,1){$\scriptstyle a_k$}
\psline[linewidth=1.5pt]{*-*}(5,-1)(5,1)
\psellipse[](8.5,0)(1.5,0.7)
\psdots*[dotstyle=triangle*,dotangle=-90,dotscale=1.2 3](8.5,.7)
\rput[b](9.5,.7){$\scriptstyle a_{k+1}$}
\psline[linewidth=1.5pt]{*-*}(7.5,0)(9.5,0)
\rput(10.5,0){$\dots$}
\psellipse[](12.5,0)(1.5,0.7)
\psdots*[dotstyle=triangle*,dotangle=-90,dotscale=1.2 3](12.5,.7)
\rput[b](13.5,.7){$\scriptstyle a_g$}
\psline[linewidth=1.5pt]{*-*}(11.5,0)(13.5,0)
\pscurve[showpoints=false,linewidth=1pt]{-}(0,.5)(1,.5)(2,.3)
\psdots*[dotstyle=triangle*,dotangle=85,dotscale=1.2 3](1,.5)
\rput[bl](.9,.6){$\scriptstyle b_1$}
\pscurve[showpoints=false,linewidth=1pt,
linestyle=dashed]{-}(2,.3)(2.5,.2)(3,-1)(1,-2)(-1,-1)(-.5,.4)(0,.5)
\pscurve[showpoints=false,linewidth=1pt]{-}(0,.8)(2,1.8)(5,.7)
\psdots*[dotstyle=triangle*,dotangle=85,dotscale=1.2 3](2,1.8)
\rput[b](3.4,1.6){$\scriptstyle b_k$}
\pscurve[showpoints=false,linewidth=1pt,
linestyle=dashed]{-}(5,.7)(5.5,.3)(6,-1.3)(1,-2.5)(-1,-1.5)(-.6,.6)(0,.8)
\pscurve[showpoints=false,linewidth=1pt]{-}(0,.9)(3,2.5)(6,2.2)(8.5,0)
\rput[b](6.2,2.3){$\scriptstyle b_{k+1}$}
\psdots*[dotstyle=triangle*,dotangle=75,dotscale=1.2 3](6,2.2)
\pscurve[showpoints=false,linewidth=1pt,
linestyle=dashed]{-}(8.5,0)(8,-3)(-1.5,-2.7)(-1.5,.7)(0,.9)
\pscurve[showpoints=false,linewidth=1pt]{-}(0,.95)(4,3.6)(10,3.2)(12,0)
\rput[b](10.2,3.3){$\scriptstyle b_g$}
\psdots*[dotstyle=triangle*,dotangle=73,dotscale=1.2 3](10,3.2)
\pscurve[showpoints=false,linewidth=1pt,
linestyle=dashed]{-}(12,0)(11.5,-3.5)(-1.9,-3.4)(-1.7,.75)(0,.95)
\endpspicture
\vspace*{2.5cm}
\end{center}
\caption{The homology basis for ${\cal L}_H$.}\label{fig1}
\end{figure}

\noindent
Let $\{\diff\omega_i\}$ denote a basis of $H^1({{\cal L}_H})$ such that
$\oint_{a_i}\diff\omega_j=2\pi{\rm i}\delta_{ij}$. Using these 
normalized differentials, we define the Abel--Jacobi map of $P_H\in{\cal L}_H$
by
$\omega(P_H)=(\int_{P_0}^{P_H}\diff\omega_1,\dots,\int_{P_0}^{P_H}\diff\omega_g)$ with
$P_0\in{\cal L}_H$ fixed. In the following, we will always choose 
$P_0=-{\rm i}z$.

We define a $g\times g$ matrix $\varPi$ -- the Riemann matrix -- with 
elements $
\pi_{ij} \doteq\oint\limits_{b_i}\diff\omega_j$.
This matrix is symmetric and has a negative definite real part. These
properties ensure that the theta function with integer characteristic
$\left [\alpha\atop\beta\right]$ defined by
\begin{equation}\label{4.2}
\Theta\left [\alpha\atop\beta\right](x,\varPi)=
\sum\limits_{N\in\SBbb{Z}^g}\exp\left\{
\frac{1}{2}\left\langle\varPi\left(N+\frac{\alpha}{2}\right),N+\frac{\alpha}{2}
\right\rangle+
\left\langle x+\pi{\rm i}\beta,N+\frac{\alpha}{2}
\right\rangle\right\}
\enspace,
\end{equation}
with $x\in{\C}^g$ and $\alpha$, $\beta\in{\Z}^g$, where $\left
\langle\cdot,\cdot\right\rangle$ denotes the Euclidean scalar product
$\left\langle N,x\right\rangle=\sum_{i=1}^gN_ix_i$, is an analytic function on
$\C^g$. A characteristic is called 
odd if $\langle \alpha,\beta\rangle \neq 0 \mbox{ mod } 2$. The Riemann 
vector is denoted by $K_R$.

The reality condition on the branch points implies for the theta function 
$\Theta$ with characteristic $\left[0\atop0\right]$, the Riemann theta 
function, 
\begin{equation}
        \bar{\Theta}(x)=\Theta(\bar{x}+
        {\rm i}\pi\Delta)\enspace,
	\label{real}
\end{equation} 
where $\Delta_i=1$ if $E_i$ and $F_i$ are real and $\Delta_i=0$ otherwise.

We recall that a divisor $\A$ on a general Riemann surface $\Sigma_g$ is a
formal
symbol $\A=n_1P_1+\cdots n_kP_k$ with $P_i\in\Sigma_g$ and $n_i\in\Z$.
The set of divisors admits a partial ordering and we may associate to a
meromorphic
function $f$ a divisor $(f)$, a principal divisor. Using this partial ordering
we define for a divisor $\A$ a vector space $L(\A)$ consisting of all principal
divisors not less than $\A$.

A Riemann--Hilbert problem can be stated as follows:
let $\Gamma$ be a piecewise smooth contour on ${{\cal L}_H}$.
Let $\Lambda=t_1+\cdots+t_r$ be a divisor
on $\Gamma$ consisting of a finite number of pairwise different points
subject to the following condition: 
$\Gamma\setminus\Lambda$ decomposes into a finite set of connected components
$\{\Gamma_j\}$ ($j=1,\dots,N$), each of which is homeomorphic to the interval
$(0,1)$. We call $\Gamma_j$ a {\em curve of the contour} $\Gamma$.
Each $\Gamma_j$ has a starting and an end point, given by two points of $\Lambda$,
where the starting respectively end points may also coincide. We define the
function $\alpha(t,\Gamma_j)$ on $\Gamma$ by
\begin{equation}\label{4.3}
\alpha(t,\Gamma_j)=\left\{
\begin{array}{ll}
1 & \mbox{if $t\in\Gamma_j$}\\
0 & \mbox{otherwise}
\end{array}\enspace,
\right.
\end{equation}
($j=1,\dots,N$). On each curve $\Gamma_j$ let there be defined a
H\"older-continuous function $G_j(t)$, which is finite and nonzero. We denote
\begin{equation}\label{4.4}
G(t)=\sum\limits_{j=1}^N\,\alpha(t,\Gamma_j)G_j(t)\enspace, \quad
t\in\Gamma\setminus\Lambda.
\end{equation}

Let there be given a divisor $\A$ of degree $m$, consisting of points of the
divisor $\Lambda$, taken at arbitrary degree. Let on ${{\cal L}_H}\setminus
\Gamma$ be given another divisor $\eufm{B}$ of degree $n$. Now we can
formulate the\\ 
{\bf Homogeneous scalar Riemann--Hilbert problem:}\\ Give a function $\psi$ 
with the properties
\begin{equation}\label{4.5}
\psi^+(t)=G(t)\psi^-(t)\enspace,
\end{equation}
with $(\psi)\in L(\A^{-1}\eufm{B}^{-1})$.

The solution of this problem was given by Zverovich~\cite{zverovich1}. A key
point in the construction is the introduction of an analogue to the usual
Cauchy kernel. This Cauchy analogue is given by a normalized (i.e.~all
$a$-periods are zero)
differential of the third kind with poles at $P_H$ and $P_0$ and is denoted by
$\diff\omega_{P_HP_0}(\tau)$. With the Cauchy analogue at hand the
solution to the above Riemann-Hilbert problem is given by
\begin{equation}\label{4.10}
\psi(P_H)={\rm e}^{\Psi(P_H)}\enspace,
\end{equation}
with
\begin{equation}\label{4.10a}
\Psi(P_H)=\frac{1}{2\pi{\rm i}}\int\limits_{\Gamma}\,\ln G(\tau)\,\diff
\omega_{P_HP_0}(\tau)\enspace
\end{equation}
where $P_0\notin\Gamma$, and the integration goes over all curves
$\Gamma_j$ of the contour $\Gamma$, and where we have put 
$\ln G(t)=\sum\limits_{j=1}^N\alpha(t,\Gamma_j)\ln G_j(t)$.
The $b$-periods $u_i$ of $\psi$ are given by
\begin{equation}
u_i=\frac{1}{2\pi {\rm i}}\int_{\Gamma}^{}\ln G\,\diff\omega_i
\label{bperiod2}\enspace.
\end{equation}
Applying the Plemelj formulae to the function $\Psi(P_H)$, we find
\begin{equation}\label{4.12}
\psi^+(t) =  G(t)\psi^-(t)\enspace,\enspace\enspace t\in\Gamma\setminus\Lambda
\setminus\bigcup\limits_{i=1}^g a_i.
\end{equation}
For more details on the solution to the scalar Riemann--Hilbert problem, 
the reader is referred to \cite{zverovich1}, \cite{rodin} or \cite{olaf}.

\subsection{Solutions to the Ernst equation via the scalar Riemann--Hilbert 
problem}

With the help of the results of the previous section,
we are now able to state the following
theorem, which gives the solution to the generalized Riemann--Hilbert problem
on the hyperelliptic Riemann surface ${\cal L}_H$.
\begin{theorem}
Let $P_0$  be a fixed complex constant ($\rho\neq0$) 
not coinciding with the 
singularities of $\psi$ or the branch points $E_i$ or $F_i$. 
Let $\Omega(P_H)$ be a linear combination of 
normalized  Abelian integrals of the 
second kind (with singularities $p\neq E_i$ and $p\neq F_i$, independent of $z$ 
and $\bar{z}$) and third kind (with in addition singularities at all real 
branch points with residues $\pm \frac{1}{2}$), satisfying 
$\bar{\Omega}(P_H)=\Omega(\bar{P}_H)$. Then the solution to the generalized
Riemann--Hilbert problem  on the real hyperelliptic Riemann surface ${\cal
L}_H$ is given by
\begin{equation}\label{8.9}
\psi(P_H)=\psi_0\,\frac{\Theta(\omega(P_H)-
\omega(D)+u+b-K_R)}{\Theta(\omega(P_H)-\omega(D)-K_R)}
\exp\left\{\Omega(P_H)+\frac{1}{2\pi{\rm i}}\int\limits_\Gamma
\ln G(\tau)\diff\omega_{P_HP_0}(\tau)\right\}\enspace,
\end{equation}
where $D=P_1+\cdots+P_g$ is a fixed non--special divisor on ${\cal L}_H$ 
which is subject to the reality condition: 
either $P_i\in \R$ or with $P_i\in D$ we have $\bar{P}_i\in D$ or $P_i$ is a 
branch point $E_i$ or $F_i$. $b$ is the vector of $b$-periods of $\Omega$ with
components
\begin{equation}
b_i=\oint\limits_{b_i}\,\Omega\label{8.10a}\enspace,
\end{equation}
$i=1,\dots,g$, the $u_i$ are given by (\ref{bperiod2}), 
and $\psi_0$ is a normalization constant. The paths of integration 
have to be the same for all integrals.
\end{theorem}
\begin{proof}
We want to prove that $\psi(P_H)$ is a single valued function on ${\cal L}_H$.
If we choose a different path of integration for the integrals in the
exponent and the map $\omega(P_H)$ and denote the corresponding integrals by a 
prime, the primed and unprimed integrals are connected via 
\begin{equation}\label{8.11}
\Omega'(P_H)=\Omega(P_H)+\oint\limits_{{\cal E}}\,\diff\Omega\enspace,
\end{equation}
(similarly for the other integrals) where ${\cal E}$ is a closed contour on 
${\cal L}_H$ which may be decomposed in the homology basis as follows
\begin{equation}\label{8.12}
{\cal E}=\sum_{i=1}^g m_i\,a_i+\sum_{i=1}^g n_i\,b_i\enspace,
\end{equation}
with $m_i$, $n_i\in\Z$. Then we have, e.g.~for $\Omega$ and $\omega$
\begin{eqnarray}\label{8.13}
\Omega(P_H) & \to & \Omega(P_H)+\sum_{i=1}^gn_i b_i=\Omega(P_H)+
\left\langle N,b\right\rangle\enspace,\nonumber\\
\omega(P_H) & \to & \omega(P_H)+2\pi{\rm i}M +\varPi N\enspace,
\end{eqnarray}
where $M=(m_1,\dots,m_g)$, $N=(n_1,\dots,n_g)\in{\Z}^g$. Under this 
transformation, the original quotient of theta functions in (\ref{8.9}) 
will be multiplied by 
\begin{equation}\label{8.14}
\exp\left (-\left\langle N,b\right\rangle\right )\enspace,
\end{equation}
but this term is just compensated by the contour integral over ${\cal E}$ in the
exponent.
The same argument holds for the line integral over the contour $\Gamma$ 
since the $u_i$ are its $b$-periods.
This shows that $\psi(P_H)$ is a single valued function on ${\cal L}_H$. 

>From the properties of the theta function, we also find that $\psi(P)$ has
$g$ simple poles at the points $P_1,\dots,P_g$ and $g$ simple 
zeros. Additional poles, zeros and essential singularities can be obtained by a
suitable choice of
Abelian integrals of the second kind (essential singularities) 
and third kind (zeros
and poles). We remark that the assumption $\bar{\Omega}(P)=\Omega(\bar{P})$ had
to be introduced in order to satisfy the reality condition of Theorem 2.1. 
\end{proof}
\begin{remark}
Without loss of generality we can choose $D$ to consist only of branch
points since  $D$ gives the poles of $\Psi$ due to the zeros of the theta
function in the denominator. This can always
be compensated by a suitable choice of the zeros and poles of $\Psi$
which arise from the integrals of the third kind in
$\Omega$. All $P_i\in D$ shall have multiplicity 1 and be chosen in a way
that $\Theta\left[\alpha\atop\beta\right](x)$ with
$\left[\alpha\atop\beta\right] =\omega(D)+\omega(\bar{P}_0)+K_R$ 
has the same reality
properties as the Riemann theta function $\Theta(x)$.
\end{remark}

Our next aim is to define a matrix
valued function $\Phi(P)$ on ${\cal L}$, satisfying the conditions of
theorem~\ref{theorem2.1}, with the help of the above solution to the scalar
Riemann--Hilbert problem on the hyperelliptic surface ${\cal L}_H$. 
To this end we define a further function on ${\cal L}_H$ by
\begin{equation}
\chi(P_H)=\chi_0\frac{\Theta(\omega(P_H)+u-\omega(\bar{P}_0)
-\omega(D)-K_R)}{\Theta(\omega(P_H)-\omega(D)-K_R)}\exp\left(
\frac{1}{2\pi {\rm i}}\int_{\Gamma}^{}\ln G\diff\omega_{P_H P_0}\right)
	\label{gauge17},
\end{equation}
where $\chi_0$ is again a normalization constant. 
It can be easily seen that the analytic behaviour of $\chi(P_H)$ is
identical to that of $\psi(P_H)$, except that it changes the sign at every
$a$-cut. $\chi$ is thus not a single valued function on ${\cal L}_H$.
However, it is single valued on $\hat{{\cal L}}$ which can be
viewed as two copies of ${\cal L}_H$ cut along 
$\left[P_0,\bar{P}_0\right]$ and glued together along this cut. We define 
the vector $X$  on $\hat{{\cal L}}$ by fixing the sign in front of $\chi$ in the 
vicinity of the points $P_0^{\pm}=(K_0,0,\pm\hat{\mu}(K_0))\in \hat{{\cal L}}$,
\begin{equation}
	X(\hat{P})=\left(
	\begin{array}{c}
		\psi(\hat{P})  \\
		\pm \chi(\hat{P})
	\end{array}\right), \quad \hat{P}\sim P_0^{\pm}.
	\label{gauge18}
\end{equation}
With the help of this vector, we can construct the matrix $\Phi$ on ${\cal 
L}$ via
\begin{equation}
	\Phi(P)=(X(K,\mu_0(K),+\hat{\mu}(K)),X(K,\mu_0(K),-\hat{\mu}(K)))
	\label{gauge19}
\end{equation}
where the signs are again fixed in the vicinity of $P_0^{\pm}$. Notice 
that this matrix consists of eigenvectors of the monodromy matrix, 
$LX(K,\mu_0(K),\pm\hat{\mu}(K))=\hat{\mu} X(K,\mu_0(K),\pm\hat{\mu}(K))$ 
if $L$ is written as $L=\hat{\mu}\Phi \gamma 
\Phi^{-1}$ where $\gamma$ is the matrix from~\ref{theorem2.1}.

It may be readily checked that this ansatz is in accordance with the reduction
condition (\ref{lin7}) (this is in fact the reason why one has to define 
the function $\chi$ in the way (\ref{gauge17})). The behaviour at the
singularities is as required in condition II: For the contour $\Gamma$ and the
singularities of the Abelian integrals $\Omega$, this is obvious. At the
branch points $E_i$ and $F_i$, one gets the following behaviour: at  
points $P_i$ of the divisor $D$, the components of $\Phi$ have a simple pole, and 
the determinant diverges as $(K-P_i)^{-\frac{1}{2}}$, if this branch point 
is not a singularity of an integral of the third kind in $\Omega$ or lies 
on the contour $\Gamma$. If the same condition holds  at 
the remaining branch points, the components are regular there but the 
determinant vanishes as $(K-P_i)^{\frac{1}{2}}$. If the branch points 
coincide with one of the singularities of the integrals in the exponent in
(\ref{8.9}), this merely changes the singular behaviour of $\Phi$ and its 
determinant there. Condition II of theorem~\ref{theorem2.1} is however 
obviously satisfied.

Since $\Phi$ in (\ref{gauge19}) is only a function of $P$, it will not be
regular at the cuts $\left[E_i,F_i\right]$. At the $a$-cuts around non-real
branch points, we get $\Phi^-=\Phi^+\sigma_1|_{a_i}$, whereas we have
$\Phi^-|_{a_i}=-\Phi^+|_{a_i}\sigma_2$ at
the $a$-cuts around real branch points. The logarithmic derivatives of $\Phi$
with respect to $z$ and 
$\bar{z}$ are however holomorphic at all these points. One can recognize that 
the behaviour at the non-real branch
points is related to a gauge transformation of the form (\ref{c}). This 
means that one can find a gauge transformed matrix $\Phi'$ that is 
completely regular at these points if the integrals in the exponent are 
regular there. With 
\begin{equation}
	\alpha_1=\frac{1}{2}(1+\lambda), \quad \alpha_2=\frac{1}{2}(1-\lambda)
	\label{cc}
\end{equation} 
and $\lambda=\prod_{i=1}^{g}\sqrt{\frac{K-\bar{P}_i}{K-P_i}}$ where
$D=\sum_{i=1}^{g}P_i$, this may be checked by direct calculation. The real
branch points, however, cannot be related to gauge transformations.

Normalizing $\psi$ and $\chi$ (if possible)
in a way that $\psi(\infty^-_H)=1$ and $\chi(\infty^-_H)=-1$, one can see 
that $\Phi$ is then in accordance with all conditions of Theorem 1 since 
the reality condition follows from the reality properties of the theta 
functions and the Riemann--Hilbert problem. The fact that $\Phi$ is at least
differentiable with respect to $z$ and $\bar{z}$ at points where $P_0$ 
does not coincide with the singularities of the integrals in the exponent 
or the remaining branch points of ${\cal L}_H$ follows from the modular 
properties of the theta function. Let the paths between 
$\left[P_0,\infty^-\right]$ and 
$\left[P_0,\infty^+\right]$ be the same in 
all integrals and let them have the same projection into the complex plane 
(i.e.\ one is the involuted of the other).
Then the results may be summarized in
\begin{theorem}\label{theorem3.2}
Let $\Theta\left[\alpha\atop\beta\right](\omega(\infty^-)+u)\neq0$. Then the function	
	\begin{equation}\label{8.16}
	f(z,\bar{z})=\frac{\Theta\left[\alpha
	\atop\beta\right](\omega(\infty^{+})+u+b)}{\Theta\left[\alpha
	\atop\beta\right]
	(\omega(\infty^{-})+u+b)}
	\exp\left\{\Omega(\infty^{+})-\Omega(\infty^{-})+
	\frac{1}{2\pi{\rm i}}\int\limits_\Gamma
	\ln G(\tau)\diff\omega_{\infty^{+}\infty^-}(\tau)
	\right\}
	\end{equation}
	is a solution to the Ernst equation. 
\end{theorem}
\begin{remarks}
\item In the case $g=0$ the Ernst potential (\ref{8.16}) is real, $f={\rm
e}^{2U}$. This means
that $U$ is a solution to the axisymmetric Laplace equation and belongs 
therefore to the
Weyl-class. For $g>0$, there are no real solutions other than $f=1$ which
describes Minkowski space.
\item The multi-black-hole solutions which can be obtained via B\"acklund
transformations (see e.g.\ \cite{hkx}) are contained in the class 
(\ref{8.16}) as the limiting case that the branch points $E_i$ and $F_i$ 
coincide pairwise. In this limit, all branch points become double points 
and the theta functions break down to purely algebraic functions. Notice 
that the analysis of $f$ at the branch points in the following section 
always assumes a regular surface. The obtained results for the regularity 
of $f$ do not hold in this limit. 
\end{remarks}
The above explicit construction of the solutions makes it possible to
derive useful formulae for the metric function $a$ and the derivatives of the
Ernst potential. Let $\int_{P_H}^{P_H+\delta}\diff\omega_i= 
g_i\delta+o(\delta)$ where
$\delta$ is the local parameter in the vicinity of $P_H\in {\cal L}_H$. We
define the derivative
\begin{equation}
	D_{P_H}\Theta(x)=\sum_{i=1}^{g} g_i\partial_{x_i}\Theta(x).
	\label{local}
\end{equation}
Using (\ref{a}) and (\ref{8.9}), (\ref{gauge17}), we get 
\begin{equation}
	(a-a_0) {\rm e}^{2U}={\rm i}D_{\infty^-}\ln 
	\frac{\Theta\left[\alpha
	\atop\beta\right]\left(\int_{\bar{P}_0}^{\infty^- }d\omega+u+b\right)}{
	\Theta\left[\alpha
	\atop\beta\right]\left(\int_{P_0}^{\infty^-}d\omega+u+b\right)}
	\label{fay6}.
\end{equation}
>From the linear system (\ref{lin1}) and (\ref{lin2}), we obtain with
(\ref{8.9}) and (\ref{gauge17})
\begin{eqnarray}
	\frac{\bar{f}_z}{f+\bar{f}}&=&\frac{{\rm i}}{2\sqrt{P_0-\bar{P}_0}}
	\frac{\Theta\left[\alpha
	\atop\beta\right](u+b-\omega(\bar{P}_0))\Theta\left[\alpha
	\atop\beta\right](u+b+
	\omega(\infty^-))}{\Theta\left[\alpha
	\atop\beta\right](u+b)\Theta\left[\alpha
	\atop\beta\right](u+b+
	\omega(\infty^-)-\omega(\bar{P}_0))}\times\nonumber\\
	&&\left(D_{P_0}\ln \Theta\left[\alpha
	\atop\beta\right](u-\omega(\bar{P}_0))+I_{P_0}\right)
	\label{fay7}
\end{eqnarray}
and 
\begin{eqnarray}
	\frac{f_z}{f+\bar{f}}&=&\frac{{\rm i}}{2\sqrt{P_0-\bar{P}_0}}
        \frac{\Theta\left[\alpha
	\atop\beta\right](u+b)\Theta\left[\alpha
	\atop\beta\right](u+b+
	\omega(\infty^-)-\omega(\bar{P}_0))}{\Theta\left[\alpha
	\atop\beta\right](u+b-\omega(\bar{P}_0))\Theta\left[\alpha
	\atop\beta\right](u+b+
	\omega(\infty^-))}\times\nonumber\\
	&&\left(D_{P_0}\ln \Theta\left[\alpha
	\atop\beta\right](u-\omega(\bar{P}_0))+I_{P_0}\right)
	\label{fay8},
\end{eqnarray}
where $I_{P_0}$ is the linear term of the expansion of the integrals 
in the exponent of (\ref{8.9}) in the local parameter around $P_0$.

\subsection{Finite gap solutions and Picard--Fuchs equations}

The original finite gap solutions of~\cite{korot1} are those among
(\ref{8.16}) without the contour integral (in our notation
only an arbitrary linear combination of Abelian integrals of the second and
third kind $\Omega$). They just correspond to the so called
Baker--Akhiezer function (see \cite{algebro}) for the Ernst system. This
function that has essential singularities and poles gives the periodic or
quasiperiodic solutions to the integrable nonlinear evolution equations. 
There the essential singularity is uniquely determined by the structure of 
the differential equation. In contrast to these equations, the solutions 
(\ref{8.16}) are in general neither periodic nor quasiperiodic, and the 
essential singularity can be nearly arbitrarily chosen. The form of the 
solution to the Riemann--Hilbert problem shows that one might even think
of ``putting the singularities densely on a line and integrate over the
integrals with some measure": an Abelian integral $\Omega_p$ of the second
kind with a pole of first order at $p$ can be used as an analogue to the Cauchy
kernel. A contour integral over this kernel with some measure,
$\int_{\Gamma}^{}\ln G \Omega_p\diff p$, is thus
just another way to write down the solution to a Riemann--Hilbert problem 
on a Riemann surface. 

In studying the boundary value problem for the rigidly rotating disk of dust,
Meinel and Neugebauer~\cite{meinelneugebauer} observed that it is possible to
obtain solutions to the Ernst equations via
\begin{equation}
f=\exp\left(\sum_{m=1}^{g}\int_{E_m}^{C_m}\frac{K^g\diff K}{\mu_H}-I_g\right)
\label{mn1}
\end{equation}
where the divisor $C=\sum_{m=1}^{g}C_m$ is determined by
\begin{equation}
	\sum_{m=1}^{g}\int_{E_m}^{C_m}\frac{K^i \diff K}{\mu_H}=I_i
	\label{mn2}
\end{equation}
($i=0,1,\dots,g-1$), i.e.\ as the solution of a Jacobi inversion problem. The
$I_i$ are (in the absence of real branch points)
real solutions to the axisymmetric Laplace equation which satisfy the 
recursive condition,
\begin{equation}
        {\rm i} I_{n+1,z}=zI_{n,z}+\frac{1}{2}I_n\enspace.
        \label{8.18}
\end{equation}
The relation to the class obtained in theorem~\ref{theorem3.2} is the following:
The integral of the third kind in
(\ref{mn1}) can be expressed by the help of a formula in~\cite{stahl}
via theta functions. Equation (\ref{mn2}) ensures that the resulting 
expression is independent of the chosen integration path which is shown in 
the proof of the theorem. Thus the $u_i$ (obtained from the $I_i$ by 
normalization) are as in our case the $b$-periods
of the integral $I_g$ in the exponent. In fact, it was shown in
\cite{meinelneugebauer} that one of these periods, say $I_1$, can be chosen 
as an arbitrary solution to the axisymmetric Laplace equation. The other 
periods as well as the integral in the exponent then follow from 
differential identities plus boundary conditions.

The underlying reason for this fact is that the 
Ernst potential $f$ is studied on a family of Riemann surfaces parametrized
by the moving branch points $-{\rm i}z$ and ${\rm i}\bar{z}$. The periods 
on this surface (i.e.~integrals along closed curves) are
subject to differential identities, the so called Picard--Fuchs equations. 
It is a general feature of the periods of rational functions 
\cite{griffiths,morrisson,foucault} that 
they satisfy a differential system of finite order with Fuchsian 
singularities. An elegant way to find the Picard--Fuchs system explicitly is
via the notion of the Manin connection in the bundle $H^1_{\rm
DR}(\Sigma_g)\to\Sigma_g$, see~\cite{manin1}. The investigation turns out to
be particularly simple if one uses the following standard form of the
(hyperelliptic) Riemann surface $\Sigma_g$ (all hyperelliptic surfaces of 
genus $g$ are conformally equivalent to this standard form)
\begin{equation}
y^2=(x-z)\prod\limits_{i=1}^{2g} (x-E_i)\doteq (x-z)P(x)
=(x-z)\sum_{j=0}^{2g}a_jx^j\enspace,
\end{equation}
where the $E_i$ do not depend on $z$. Using $j_0=\diff x/y$, 
$j_1=xj_0,\dots,j_{2g-1}=x^{2g-1}j_0$ as the basis for the de Rham 
cohomology $H^1_{\rm DR}(\Sigma_g)$ we
obtain for the matrix $M_n^m$ ($m,n=0,\dots,2g-1$) of the Manin connection
(defined by $\frac{\partial j_n}{\partial z}=M^m_nj_m$)
\begin{equation}
M_n^m=
\left\{
\renewcommand{\arraystretch}{1.4}
\begin{array}{cl}
\displaystyle \frac{z^n}{2P(z)}\left((m+1)a_{m+1}+
z^{-m-1}\,\sum_{j=0}^ma_j z^j\right)& \mbox{for $0\leq m<n$,} \\
\displaystyle \frac{z^n}{2P(z)}\left((m+1)a_{m+1}-
\sum_{j=0}^{2g-1-m}a_{m+1+j}z^j\right) & \mbox{for $n\leq m\leq 2g-1$} \\
\end{array}\right.\enspace.
\end{equation}
One finds that the periods satisfy a similar recursive condition as
(\ref{8.18}).
An analogous consideration can be performed for the $\bar{z}$-dependence of
the periods. One finds that the integrability condition of the Picard--Fuchs
systems is just the axisymmetric Laplace equation.

On the other hand, with the help of some boundary conditions (for instance at
$|z|\to\infty$), the $I_n$ can be uniquely determined from the above system
(\ref{8.18}).
Thus the class of solutions discussed by Meinel and Neugebauer may be phrased
in the following form: if an arbitrary solution of the Laplace equation is
given, one can calculate the functions $I_n$ with (\ref{8.18}) and the boundary
condition, and ends up with a solution to the Ernst equation of the form 
(\ref{8.16}).

\section{The singular structure of the Ernst potential}
\setcounter{equation}{0}

The construction of the solutions  in the previous sections with 
the help of Theorem~\ref{theorem2.1} also indicates where the resulting Ernst
potential
(\ref{8.16}) may be singular: only at points $P_0=-{\rm i}z$ where the 
conditions of Theorem~\ref{theorem2.1} do not hold. Notice that these
conditions are sufficient for the regularity of $f$ at all other points. It may
turn out though that the Ernst potential is perfectly regular at points where
Theorem~\ref{theorem2.1} is not fulfilled, e.g.\ in the case of singularities
that are pure gauge. We will therefore discuss all possible singular points of
the solutions (\ref{8.16}).

It is very helpful that this whole discussion can be  performed on the 
Riemann surface ${\cal L}_H$ where one can use the powerful calculus on
hyperelliptic surfaces.
One does not have to work on the four sheeted surface $\hat{{\cal L}}$ 
whose introduction was necessary for the
construction of the solutions, and which provides an understanding
of the mathematical properties of the Ernst equation. Since we will work 
from now on on ${\cal L}_H$ only, unless otherwise noted, we will drop the 
index $H$ at points $P\in {\cal L}_H$.

The possible singularities of $f$ can be directly inferred 
from the potential in the form (\ref{8.16}). The Ernst potential will be 
singular at the zeros of the denominator.
It is possibly not regular at the points 
where $P_0$  is identical to the singularities of $\Omega$ or
is on $\Gamma$. Critical points of a 
different kind are the branch points $E_i$ and $F_i$. If $P_0$  
coincides with these points, the Riemann surface ${\cal L}_H$
becomes singular. Something similar happens at the axis where the 
branch points $P_0$ and $\bar{P}_0$ coincide. This is a reminiscent of the
singular behaviour of the three--dimensional Laplace operator on the axis 
in the axisymmetric case. The
main aim of the following analysis is to single out a class of solutions that 
may be interesting in the context of boundary value problems for the Ernst 
equation that describe e.g.\ the exterior of a body of revolution. Thus we 
will not study the nature of the singularities (e.g.\ curvature 
singularities) but single out a large class of solutions where the Ernst 
potential is only discontinuous at a (closed) contour that could be identified 
with the surface of a body.

\paragraph{Zeros of the denominator}

Zeros of the denominator of (\ref{8.16}) will lead to singularities in the 
spacetime. From condition IV of Theorem 2.1 it follows that these are just 
the points at which the matrix $\Phi$ cannot be normalized in the required 
way. This leads to the transcendental condition
\begin{equation}
	\Theta\left[\alpha\atop\beta\right](\omega(\infty^-)+u+b)\neq 0,
	\label{reg}
\end{equation}
if one wants to exclude these zeros of the denominator.
We will show in the next section 
how the points at which 
$\Theta\left[\alpha\atop\beta\right](\omega(\infty^-)+u+b) =0$  
can be found as the solution of a set of algebraic equations.

\paragraph{Essential singularities}

The integrals of the third kind occuring in $\Omega$ are nothing but
a particular case of line integrals over contours with constant jump function
$G(t)$. Therefore, we are left with the investigation of the integrals of the
second kind at this point. Since the theta functions in (\ref{8.16}) are 
regular as long as the Riemann surface ${\cal L}_H$ is, we are left with 
the exponent if (\ref{reg}) holds. 
For the behaviour of the exponent, we get the following
\begin{proposition}\label{prop4.1}
The Ernst potential (\ref{8.16}) has an essential singularity at the points
where $P_0$  coincides with the singularities of the integrals of
the second kind on ${\cal L}_H$.
\end{proposition}
\begin{proof}
The exponent has by construction an algebraic
pole there and, therefore, the Ernst potential has an essential
singularity.
\end{proof}
An essential singularity of the real part of the Ernst potential 
corresponds to a line singularity of the metric function $U$. 
In the context of exterior solutions for bodies of revolution we are 
interested in, there seems to be no 
situation where such a line singularity in a spacetime might be interesting.

\paragraph{Contours}

In the case that $P_0$ lies on a contour $\Gamma_i$ but
not on an endpoint of $\Gamma_i$,
on the axis, or on one of the branch points $E_i$ or $F_i$, it
can be easily seen that the
integral in the exponent as well as the $b$--periods $u_i$ are bounded 
since $G$ is H\"older--continuous, finite and non--zero on $\Gamma$. At 
the endpoints of the contours $\Gamma_i$, singularities may occur.
The value of these integrals at the remaining points will 
however not be the same in general if the contour is approached from one or 
the other side. This can be seen from the following fact: The point 
$P_0$ is a branch point of ${\cal L}_H$. If it lies on the contour
$\Gamma$, care has to be taken of the sign of the root whilst evaluating 
the integrals of the form $J_n=\int_{\Gamma}^{}\ln G(\tau)\frac{ \tau^n 
\diff \tau}{\mu}$. The decisive factor is $\mu_0^2=(K-P_0)(K-\bar{P}_0)$.  
We get for $K\in\Gamma$ with $K=K_1+{\rm i}K_2$,  and $K_1,K_2\in \R$ 
for the imaginary part of $\mu_0$,
\begin{equation}
	\Im\mu_0=\pm\frac{1}{\sqrt{2}}\mbox{sgn}\left((K_1-\zeta)K_2\right)
	\sqrt{|\mu_0^2|-\Re\left(\mu_0^2\right)}
	\label{4.2a},
\end{equation}
i.e.\ the sign of the imaginary part of 
$\mu$ depends on the sign of $K_1-\zeta$. Thus the value of the integrals 
will in general not be the same whether the contour is approached from the 
interior or the exterior region.
This reasoning does not work for points $K=K_0$ 
not on the axis ($K_2\neq 0$) with $K_1=0$. There the imaginary part 
of $\mu_0^2$ is zero which means that $\mu_0$ is either purely 
imaginary or real in the vicinity of $P_0=K_0$ depending  on the sign of
$K_2-\zeta$. 
We conclude that the integrals over $\Gamma$ with $P_0\in \Gamma$ have the form 
$J=J^1+\mbox{sgn}(\epsilon) J^2$ (where the $J^i$ are independent of 
$\epsilon$ which indicates if the contour is approached from the interior 
or the exterior) which implies that the limiting value of the 
Ernst potential calculated via (\ref{8.16}) exists but depends on
$\epsilon$. Therefore, we have proven the following
\begin{proposition}
Let $P_0$ lie on the contour $\Gamma$ but not on the
axis, on one of the branch points $E_i$ or $F_i$, or at an endpoint 
of $\Gamma_i$. Then $f$  will, in general, have a jump at
$\Gamma$. The limiting value of $f$ will exist and be H\"older--continuous 
there. $f$ may be singular at the endpoints of the $\Gamma_i$.
\end{proposition}
Thus the Ernst potential will be finite but discontinuous at a 
contour $\Gamma_z$ in the $(\rho,\zeta)$--plane given by $P_0\in \Gamma$ 
which means that the 
solution to the vacuum equations will not be regular at a surface in the 
$(\rho$, $\zeta$, $\phi$)--space. If this surface is closed, it can 
possibly be 
identified with the surface of a body of revolution. The interior of the 
body is supposed to be filled with matter. Therefore the vacuum solution is 
only considered in the exterior (that contains $z=\infty$); it is not 
regular at the boundary to the matter region.

\paragraph{The axis}

The axis is a double point on the Riemann surface ${\cal L}_H$ since two branch
points coincide. In this case, all quantities may be considered on
the Riemann surface $\Sigma'$ given by 
$\mu'{}^2=\prod_{i=1}^{g}(\tau-E_i)(\tau-F_i)$, as was shown by Fay
\cite{fay}. Let a prime denote here and in the following that the primed 
quantity is taken on $\Sigma'$. This surface is obtained from ${\cal L}_H$ 
by removing the cut $\left[P_0,\bar{P}_0\right]$. For the 
analysis of the axis, we will use a slightly different cut--system than the 
one introduced in the previous section: we take a closed curve around 
$\left[P_0,\bar{P}_0\right]$ in the $+$--sheet as the cut $a_g$. All $b$--cuts 
shall begin at the cut $\left[E_1,F_1\right]$. The rest is unchanged.
This implies for the characteristic of the theta function that it has the form
\begin{equation}
	\left[
	\begin{array}{ll}
		\alpha' & 1  \\
		\beta' & \varepsilon
	\end{array}
	\right]
	\label{chara}
\end{equation}
where $\varepsilon=0,1$ and $\alpha_i'=0$. 

Since the expansions of all characteristic quantities of the Riemann surface
are smooth in $\rho$ except $\pi_{gg}$ which is divergent as $\ln \rho$ for
$\rho\to 0$, it follows that the Ernst potential has a regular expansion 
in $\rho$. For points $P_0$ not coinciding with real branch points or 
singularities of the exponent in (\ref{8.16}), the Ernst potential is thus 
at least $C^3$. It follows from a theorem of M\"uller zum Hagen \cite{mzh} that 
it is therefore analytic. Consequently it is sufficient 
to calculate the limiting case. If this is well defined, the 
Ernst potential is regular at these points of the axis. 
The differentials of the first kind for $\rho=0$ turn out to be
\begin{equation}
	\diff\omega_i=\diff\omega_i', \quad i=1,\ldots, g-1, \quad
	\diff\omega_g=-\diff\omega'_{\zeta^+\zeta^-},
	\label{sing1}
\end{equation}
where $\diff\omega'_{\zeta^+\zeta^-}$ is the normalized differential of 
the third kind on $\Sigma'$ with poles in $\zeta^+$ and $\zeta^-$. 
This implies for the $b$-periods
\begin{eqnarray}
	\pi_{ij} & = & \pi_{ij}',\quad i,j=1,\ldots, g-1,
	\label{sing2} \\
	\pi_{ig} & = & -\int_{\zeta^-}^{\zeta^+}\diff \omega_i', \quad i=1,\ldots, 
	g-1,
	\label{sing3} \\
        \pi_{gg} & = & 2\ln\rho+\mbox{reg. terms}\enspace.
	\label{sing4}
\end{eqnarray}
Since $\pi_{gg}$ diverges, the theta function will break down to a sum of 
two theta series on $\Sigma'$ (in the case of genus $g=1$, the surface 
$\Sigma'$ has genus 0; the formula below can however be used if one 
replaces the theta function $\Theta'$ simply by a factor 1 which means 
that the axis potential can be expressed in terms of elementary functions 
in this case). We introduce integrals $\omega'(P)$ of the first kind with the
property $\omega'(E_1)=0$. 
The differential $\diff\omega_{\infty^+\infty^-}$ on the axis becomes
$\diff\omega'_{\infty^+\infty^-}$. In the case of the contour integrals 
one has to observe that an additional factor $\mbox{sgn}(K_1-\zeta)$ in the 
notation of (\ref{4.2a}) occurs for the same reasons as there. Since the 
Abelian integrals of the second kind can be obtained from the integrals of 
the third kind by a limiting procedure, the same holds for these integrals 
and their $b$-periods. With the above settings, we obtain for (\ref{8.16})
\begin{eqnarray}
        f&=&\frac{\Theta'\left[\alpha'\atop\beta'\right]
        \left(\omega'|^{\infty^+}_{\zeta^+}+u'+b'\right)
	+(-1)^{\varepsilon}\exp(-(\omega'_g(\infty^+)+u_g+b_g))
	\Theta'\left[\alpha'\atop\beta'\right]\left(\omega'|^{\infty^+}_{\zeta^-}
	+u'+b'\right)}{\Theta'\left[\alpha'\atop\beta'\right]
        \left(\omega'|^{\infty^+}_{\zeta^+}-u'-b'\right)
	+(-1)^{\varepsilon}\exp(-(\omega'_g(\infty^+)-u_g-b_g))
	\Theta'\left[\alpha'\atop\beta'\right]\left(\omega'|^{\infty^+}_{\zeta^-}
	-u'-b'\right)}
        \nonumber\\
        &&\exp\left\{\Omega'|_{\infty^-}^{\infty^+}+\frac{1}{2\pi i}
        \int\limits_\Gamma
\ln G(\tau)\diff\omega'_{\infty^{+}\infty^-}(\tau)+b_g+u_g\right\}.
	\label{sing7}
\end{eqnarray}
It can be seen from the above formula that the limiting value of $f$ exists 
even if $u_g$ diverges, provided (\ref{reg}) holds ($f$ will be
H\"older-continuous if $u_g$ diverges). The Ernst potential will however have
an essential singularity at the real singularities of $\Omega$. We can
summarize the above results.
\begin{proposition}
Let condition (\ref{reg}) hold. Then the
Ernst potential is regular on the axis except at the points where $P_0$
coincides with singularities
of $\Omega$, points of $\Gamma$, and branch points $E_i$, $F_i$.
\end{proposition}
\begin{remark}
        Though $f$ is H\"older-continuous even if $u_g$ diverges, it is
        interesting to note for the following when this will be the case.
	Obviously this can only happen at the 
	real points of $\Gamma$. It can be seen however that $u_g$ is always 
	bounded at these points due to the reality condition, unless they are 
	endpoints of $\Gamma_i$ (this would lead to a conic singularity on the 
	axis).
\end{remark}

\paragraph{Real branch points}

If $P_0$ coincides with a real branch point $E_i$ or $F_i$, this will be a 
triple point on ${\cal L}_H$. We get the following.
\begin{proposition}
	At points where $P_0$ coincides with the real branch point $E_g$, the 
	limiting value of $f$ exists. The Ernst potential 
	is in general not differentiable there. 
\end{proposition}
\begin{proof}
	We use the same cut system and the same notation as on the axis. Put 
	$P_0=E_g+x$ with $x=\delta {\rm e}^{{\rm i}\phi}$ and $\phi\in\R$, 
	$\delta\in\R^+$. In order to expand $f$ in powers of $x$ and $\bar{x}$, 
        one has to consider the $a$-periods, in particular
	\begin{eqnarray}
\oint_{a_g} \frac{\diff\tau}{\mu} &=& \frac{4}{\sqrt{\bar{x}(F_g-E_g)}}
\int_{1}^{\frac{1}{k}}\frac{\diff t}{\sqrt{(1-t^2)(1-k^2 t^2)(F_g-E_g+xt^2)}
\mu''(F_g+xt^2)}\nonumber\\
&=&\frac{4}{\sqrt{\bar{x}(F_g-E_g)}\mu''(F_g)}
({\rm i}\tilde{K}(k)+O(\delta))
		\label{reala} 
\end{eqnarray}
where $k={\rm e}^{{\rm i}\phi}$, where
$\tilde{K}(k)=K(\sqrt{1-k^2})$ and $K(k)$ are the complete
elliptic integrals of the first kind, and where
$\mu''{}^2(\tau)=\prod_{i=1}^{g-1}(\tau-E_i) (\tau-F_i)$. It can be seen
from (\ref{reala}) that the $a$--period has an expansion in powers of
$\sqrt{\delta}$. The coefficients of the expansion in $\sqrt{x}$,
and $\sqrt{\bar{x}}$ are $\phi$--dependent, since the modul of the elliptic
integrals is just $k= {\rm e}^{{\rm i}\phi}$.
This implies for the differentials of the first kind
$\diff\omega_i=\diff\omega_i'+O(\sqrt{\delta})$ for
$i=1,...,g-1$, and $\diff\omega_g=\diff\omega_{g-1}$. Similarly
$\diff\omega_{\infty^+\infty^-}=\diff\omega'_{\infty^+
\infty^-}+O\left(\sqrt{\delta}\right)$. We get for the
$b$-periods,
\begin{equation}
\pi_{gg}=-2\pi\frac{K(k)}{K_1(k)}\left(1+O\left(\sqrt{\delta}\right)\right)
\label{realb}\enspace,
\end{equation}
whereas $\pi_{(g-1)g}=O(\sqrt{\delta})$ and $\pi_{ij}=\pi_{ij}'$ for
$i,j=1,\dots,g-1$ in the limit. Thus $f$ can be expanded in $\sqrt{x}$ and $
\sqrt{\bar{x}}$. Even in case that only integer powers in the expansion
occur, the coefficients will be in general $\phi$-dependent. Though the
limiting value of $f$ at $P_0=E_g$ exists, $f$ will in general not be
differentiable at this point.
\end{proof}
This implies that the real branch points are singular points on the axis,
possibly topological defects in the spacetime, see \cite{korot1}. They should not occur in 
the context of exterior solutions for bodies of revolution we are 
interested in.

\paragraph{Non-real branch points}

If $P_0$ coincides with a branch point $E_g=\bar{F}_g$, the points $E_g$ 
and $F_g$ will be double points on ${\cal L}_H$. Thus the situation is 
similar to the one on the axis with the only exception that one ends up 
here with two double points.  
As on the axis, it is convenient to consider all quantities on 
a Riemann surface $\Sigma''$ given by 
$\mu''{}^2=\prod_{i=1}^{g-1}(K-E_i)(K-F_i)$ where the double points are
removed. All quantities with two primes 
are understood to be taken on this surface. We use the following cut 
system: let $a_{g-1}$ be the circle around $\left[P_0,E_g\right]$, and $a_g$ 
the circle around $\left[\bar{P}_0,F_g\right]$, both in the plus 
sheet. The remaining cuts are as on the axis, i.e.\ all $b$--cuts start at 
$\left[E_1,F_1\right]$.  As on the axis, we get
\begin{proposition}
Let (\ref{reg}) hold, and let $E_g=\bar{F}_g\notin\Gamma$.
	Then the Ernst potential is regular at the point $P_0=E_g$. 
	For $E_g\in\Gamma$, $f$ is in general H\"older--continuous at $P_0=E_g$.
\end{proposition}
The proof is similar to the one on the axis and basically uses again results of 
Fay \cite{fay}.
\begin{proof}
	The case $g=1$ may be checked directly with the help of the standard 
        theory of elliptic theta functions (see e.g.\ \cite{erdelyi}).
	For $g>1$ with the cut system in use and $P_0=E_g+x$, 
	where $x$ is chosen as in the 
	case of the real branch points, the differentials of the 
	first kind have a smooth expansion in $x$ and $\bar{x}$. In contrast to 
	the case of real branch points, the coefficients in the expansion are 
	$\phi$--independent. The differentials
	$\diff\omega_{i}$ become in leading order the differentials of the first kind 
	$\diff\omega''_i$ on $\Sigma''$.  The differential 
	$\diff\omega_{g-1}$ becomes in the limit the differential 
	$-\diff\omega''_{E_g^+E_g^-}$, and similar for $\diff\omega_g$ at $F_g$. 
	The differential of the third kind becomes 
	$\diff\omega_{\infty^+\infty^-}=
	\diff\omega_{\infty^+\infty^-}''$. All these differentials have 
	coefficients in the $x$ and $\bar{x}$ expansion that contain Abelian 
	integrals of the second kind with poles in $E_g^{\pm}$ and $F_g^{\pm}$ as 
	may be checked by direct calculation. This implies for the 
	$b$--periods that $\pi_{ij}=\pi''_{ij}$ for $i,j=1,...,g-2$ and 
	\begin{eqnarray}
		\pi_{(g-1)(g-1)} & = & \pi_{gg}=2\ln\delta +...,
		\nonumber \\
		\pi_{i(g-1)} & = & -2\omega''(E_g^+),
		\nonumber \\
		\pi_{ig} & = & -2\omega''(F_g^+),
		\label{com1}
	\end{eqnarray} 
	whereas $\pi_{(g-1)g}$ is finite in the limit $\delta\to0$.
	If $E_g\notin \Gamma$, the $u_i$ as well as the Cauchy integral in the 
	exponent have a smooth expansion in $x$ and $\bar{x}$ with finite 
	coefficients.  The theorem of \cite{mzh} then guarantees regularity if the
	limiting value that may be calculated as on the axis exists. 
	The theta function 
	on ${\cal L}_H$ breaks down to a sum of four theta functions on 
	$\Sigma''$ times a multiplicative factor. If $E_g\in\Gamma$, 
	the coefficients in the expansion of $f$ in $x$ and $\bar{x}$ will diverge 
	which implies that $f$ is possibly 
	not differentiable there though the limiting 
	value exists if (\ref{reg}) holds.
\end{proof}

\section{Metric functions and ergospheres}

In the previous sections we have made extensive use of the 
complete integrability of the Ernst equation to 
construct a large class of solutions. To discuss physical 
features of the resulting spacetimes however, it would be helpful to have 
expressions in closed form not only for the Ernst potential but for the 
metric functions, at least for the functions ${\rm e}^{2U}$ and $a$ that 
can be expressed invariantly via the Killing vectors. It is a remarkable 
fact already noticed by Korotkin \cite{korot1} that the metric function 
$a$ can be related to derivatives of the matrix $\Phi$ without solving 
the differential equation (\ref{vac9}). In the following we will show that 
a theta identity of Fay \cite{fay} can be used to go one step 
further to obtain a formula for $a$ that is free of derivatives. The same 
identity leads to a simplified expression for the metric function ${\rm e}^{2U}$ 
that can be directly used to identify ergospheres in the spacetime.

Fay's trisecant identity establishes a relation between four points
$A_1,\dots,A_4$ on a
Riemann surface, in our case ${\cal L}_H$, in arbitrary position (see e.g.\ 
\cite{mumford}, \cite{taimanov}). Let $x$ be an arbitrary $g$-dimensional
vector. Then the following identity holds,
\begin{eqnarray}
        &  &\Theta(x)\Theta\left(x+\int_{A_1}^{A_3}\diff\omega+
        \int_{A_2}^{A_4}\diff\omega\right)
        -\exp \left(\Omega_{A_1A_4}|^{A_3}_{A_2}\right)
        \Theta\left(x+\int_{A_2}^{A_3}\diff\omega\right)
        \Theta\left(x+\int_{A_1}^{A_4}\diff\omega\right)
	\nonumber \\
        &&- \exp\left(\Omega_{A_2A_4}|^{A_3}_{A_1}\right)
        \Theta\left(x+\int_{A_1}^{A_3}\diff\omega\right)
        \Theta\left(x+\int_{A_2}^{A_4}\diff\omega\right) =  0\enspace,
	\label{fay1a}
\end{eqnarray}
where e.g.~$\Omega_{A_1A_4}|^{A_3}_{A_2}$ denotes the integral of a normalized
differential of the third kind with simple poles at $A_1$ respectively $A_4$
with residues $+1$ respectively $-1$ along a path from $A_2$ to $A_3$. For a
geometric interpretation of this identity in terms of the Kummer variety see
\cite{mumford}, \cite{taimanov}, for an interpretation via generalized cross
ratio functions see \cite{farkas}. The strength of the above identity arises
from the fact that it holds for points $A_i$ in general position. By a 
suitable choice of these points, we obtain for the metric function ${\rm 
e}^{2U}$, the real part of the Ernst potential, 
\begin{equation}
	e^{2U}=\frac{1}{2}\exp \left(\Omega_{\bar{P}_0\infty^-}|^{P_0}_{\infty^+}
	\right)
	\frac{\Theta\left[\alpha\atop\beta\right](u+b)\Theta\left[\alpha\atop\beta\right]
	(u+b+\omega(\bar{P}_0))}{\Theta\left[\alpha\atop\beta\right]
	(u+b+\omega(\infty^-)+\omega(\bar{P}_0))
	\Theta\left[\alpha\atop\beta\right](u+b+\omega(\infty^-))} e^{I}
	\label{fay14}
\end{equation}
where $I$ denotes the integral in the exponent of (\ref{8.16}).
This formula makes it possible to identify directly the zeros of ${\rm 
e}^{2U}$ which give the ergospheres, the limiting surfaces of stationarity 
(inside these surfaces there can be no observer at rest with respect to 
spatial infinity). Since the exponent of 
the integral of the third kind in (\ref{fay14}) in front of the fraction 
cannot vanish, the necessary condition for ergospheres is 
\begin{equation}
	\Theta\left[\alpha\atop\beta\right](u+b)\Theta\left[\alpha\atop\beta\right]
	(u+b+\omega(\bar{P}_0))=0
	\label{ergo}.
\end{equation}
Defining the divisor $A$ as the solution of the Jacobi inversion problem
\begin{equation}
	\omega(A)-\omega(D)=u+b
	\label{ergo1},
\end{equation}
we find that an ergosphere can occur if $P_0$ or $\bar{P}_0$ are in $A$. It
is however possible that the denominator of (\ref{fay14}) vanishes at the 
same time which would imply a violation of (\ref{reg}) (and thus a 
singularity of the spacetime). Summing up we get
\begin{proposition}
	I. Let $P_0$ or $\bar{P}_0$ and $\infty^-$ be in $A$ for some $P_0$, 
	then condition 
	(\ref{reg}) is violated and the Ernst potential is singular
	at these points.\\
	II. Let $P_0$ or $\bar{P}_0$ but not $\infty^-$ be in $A$ for some $P_0$, 
	then the real part of the Ernst potential vanishes at these points which 
	 describe an ergosphere.
\end{proposition}

The same formula can be used on the axis where we obtain for the metric 
function ${\rm e}^{2U}$ in the notation of (\ref{sing7})
\begin{eqnarray}
	{\rm 
        e}^{2U}&=&\frac{1}{2}\exp\left(
        \Omega_{\infty^-\zeta^+}|^{\infty^+}_{\zeta^-}
	\right)\nonumber\\
        &&\frac{\Theta'\left[\alpha' \atop \beta'\right]^2(u'+b')}{\Theta'
        \left
        [\alpha' \atop \beta'\right]^2(\omega'|^{\infty^-}_{\zeta^-}+u'+b')-
        \exp(2(\omega_g(\infty^-)+u_g+b_g))\Theta'\left[\alpha' \atop
        \beta'\right]^2
        (\omega'|^{\infty^-}_{\zeta^+}+u'+b')}\nonumber\\
	\label{ergo2}.
\end{eqnarray}
The condition for an ergosphere to hit the axis is then
\begin{equation}
\Theta'\left[\alpha'\atop\beta'\right](u'+b')=0\enspace,
	\label{ergo3}
\end{equation}
since the integral of the third kind in the exponent in front of the fraction
cannot diverge for finite values of $\zeta$. The interesting feature of 
this relation is that it is completely independent of the physical 
coordinates. This implies that if an ergosphere extends to the axis, this 
will be only possible if the metric function ${\rm e}^{2U}$ vanishes on the
whole axis. In the case of the Kerr solution, the ergosphere touches the 
axis at the horizon. An interpretation of the fact 
that the whole axis would be singular in the present case is given in the 
next section where the above case is related to the ultrarelativistic limit 
in which the source of the gravitational field becomes so strong that it
vanishes behind the horizon of the extreme Kerr metric.

The metric function $a$ can be calculated from (\ref{a}) if one uses the 
trisecant identity (\ref{fay1a}) in the limit that two points coincide. 
Using the trisecant identity several times, we get
\begin{eqnarray}
        (a-a_0){\rm e}^{2U}&=&-\rho\left(
        \frac{\Theta\left[\alpha\atop\beta\right](0)
        \Theta\left[\alpha\atop\beta\right]
        \left(\int_{\bar{P}_0}^{P_0}\diff\omega\right)}{
        \Theta\left[\alpha\atop\beta\right]\left(
        \int_{\infty^-}^{P_0}\diff\omega\right)
	\Theta\left[\alpha\atop\beta\right]
        \left(\int_{\bar{P}_0}^{\infty^-}\diff\omega\right)}\times
        \right.\nonumber\\
	&&\left.\frac{\Theta\left[\alpha\atop\beta\right](u+b)
	\Theta\left[\alpha\atop\beta\right]
        \left(u+b+\int_{P_0}^{\infty^-}\diff\omega+
        \int_{\bar{P}_0}^{\infty^-}\diff\omega\right)}{
	\Theta\left[\alpha\atop\beta\right]
        \left(u+b+\int_{\bar{P}_0}^{\infty^-}\diff\omega\right)
	\Theta\left[\alpha\atop\beta\right]
        \left(u+b+\int_{P_0}^{\infty^-}\diff\omega\right)}-1\right)
        \label{fay7a}\enspace.
\end{eqnarray}
The constant $a_0$ can be obtained in a similar way from the condition 
that $a=0$ on the regular part of the axis (we assume here that the 
singularities in the exponent of (\ref{8.16}) are situated in a compact 
region of the $(\rho,\zeta)$--plane). Care has to be taken in the above 
formula that some of the terms in brackets explode as $1/\rho$ in the 
limit $\rho\to0$. We get
\begin{equation}
	a_0=\frac{{\rm i}}{2}(D-\bar{D})\frac{\Theta'\left[\alpha'\atop\beta'\right]
	(u'+b'+\int_{\infty^+}^{\infty^-}d\omega')
	\Theta'\left[\alpha'\atop\beta'\right](0)}{\Theta'\left[\alpha'\atop\beta'\right]
	(\int_{\infty^+}^{\infty^-}d\omega')
	\Theta'\left[\alpha'\atop\beta'\right](u'+b')}e^{-I'}
	\label{fay18b}.
\end{equation}
It can be seen from this formula that $a_0$ does not vanish if there 
are no singularities in the exponent ($I=u=b=0$) in which case $f=1$ which 
describes Minkowski spacetime. This reflects, as already noted, the fact 
that $a_0$ is a gauge dependent quantity. The metric function $a$ however 
is gauge independent. In the above example of Minkowski spacetime, it will 
of course vanish in the used asymptotically non-rotating coordinates.

\section{Asymptotic behaviour and equatorial symmetry}

Since we are mainly interested in solutions to the Ernst equation that could
describe the gravitational field outside a compact matter source, we will study
the asymptotic behaviour (near spatial infinity) of the solutions (\ref{8.16}).
It is generally believed
that the Ernst potentials of the corresponding spacetimes are regular except
at the contour $\Gamma_z$ in 
the $(\rho,\zeta)$--plane which corresponds to the surface of the body, 
asymptotically flat and equatorially symmetric. We will investigate in the 
following whether it is possible to identify solutions with these properties in 
the class (\ref{8.16}).

\paragraph{Asymptotic behaviour}

Asymptotic flatness implies that the Ernst potential is of the form
$f=1-2m/|z|+o(1/|z|)$ for $|z| \to \infty$ where $m$ is a positive real
constant. A complex $m$ is related to a so called NUT-parameter that is
comparable to a magnetic monopole.

The asymptotic properties of the solutions (\ref{8.16}) can be read off at the 
axis. Notice that the $\diff\omega_i'$ are independent of $\zeta$. 
For $\diff\omega_g$, we get
\begin{equation}
	\diff\omega_g=\diff\omega'_{\infty^+\infty^-}\left(1-\frac{1}{2\zeta}
	\sum_{i=1}^{g}(E_i+F_i)\right)+\frac{1}{\zeta}\diff\omega'_{\infty^+,1}
	+o(1/\zeta)
	\label{sing9}
\end{equation}
where $\diff\omega'_{\infty^+,1}$ is the differential of the second kind 
with a pole of second order at $\infty^+$. Furthermore it can be seen 
that $\exp(-\omega_g(\infty^+))$ is proportional to $1/\zeta$ for $\zeta\to
\infty$. Thus we get
\begin{proposition}
        Let $\lim_{\tau\to \infty}\tau\ln G(\tau)=0$ 
        on all contours that go through
	$\infty^+$ or $\infty^-$ and let $\Theta'\left[\alpha'\atop\beta'\right]
        \left(u'+b'\right)\neq0$. Then $f$ has the form $f=1-2m/\zeta$ for 
	$\zeta\to\infty$ where $m$ is a complex constant. 
\end{proposition}
The proof of this proposition follows from (\ref{sing9}) and (\ref{sing7}).

\paragraph{Equatorial symmetry}

The fact that the mass is in general complex implies that the class we are 
considering here is too large if one wants to study only solutions that are
asymptotically flat in the strong sense ($m$ real). There is the belief that
stationary axisymmetric spactimes describing isolated bodies in thermodynamical
equilibrium are equatorially symmetric. This implies for the Ernst potential
$f(-\zeta)=\bar{f}(\zeta)$. Solutions with this property always have a real
mass due to the symmetry. It is therefore of special interest to single out
equatorially symmetric solutions among those in (\ref{8.16}). We get
\begin{theorem}
Let ${\cal L}_H$ be a hyperelliptic surface of the form (\ref{3.22a}) with 
even genus $g=2s$ and the property $\mu(-K,-\zeta)=\mu(K,\zeta)$.	
Let $\Gamma$ be a piecewise smooth contour on ${\cal L}_H$ such that 
with $P=(K,\mu(K))\in \Gamma$ also $\bar{P}\in
\Gamma$ and $(-K,\mu(K))\in \Gamma$. Let there be 
given a finite nonzero function $G$  on 
$\Gamma$ subject to $G(\bar{P})=\bar{G}(P)=G((-K,\mu(K)))$. If $(p,\mu(p))$
is a singularity of $\Omega$, the same should hold for $(-p,\mu(-p))$. 
Choose a cut sytem in a way that the cuts $a_i^1$ ($i=1,\ldots ,s$) encircle 
$\left[-F_i,-E_i\right]$ and $a_i^2$ encircle
$\left[E_i,F_i\right]$  in the $+$--sheet (in the case of real branch 
points, the points are ordered in the way $E_i<F_i<E_{i+1}<\ldots $; points 
with the same real part are ordered in the way $\Im (E_i)<\Im 
(F_i)<\Im(E_{i+1})<\ldots $ which implies that $E_i\neq \bar{F}_i$ in this 
special case).  \\
Then  $f$ is equatorially symmetric if the characteristics 
in the $i$--th position  (any combination of the two cases is 
allowed) have the form
\begin{equation}
	\left[
	\begin{array}{cc}
		0 & 0  \\
		1 & 1
	\end{array}
	\right], \quad 
	\left[
	\begin{array}{cc}
		0 & 0  \\
		0 & 0
	\end{array}
	\right]
	\label{equ1}.
\end{equation}

\end{theorem}
\begin{proof}
	The property $\mu(\zeta,K)=\mu(-\zeta,-K)$ on ${\cal L}_H$ makes it 
	possible to express quantities on a surface with $\zeta=-\zeta_0$ in terms 
	of  the 
	corresponding quantities on the surface with $\zeta=\zeta_0$. We have 
	$a_i^1(-\zeta)=\tau a_i^2(\zeta)$ and $b_i^1(-\zeta)=\tau 
	b_i^2(\zeta)$ where $\zeta$ and $-\zeta$ denote the surface on which the 
	quantity is considered, and where $\tau$ is the anti--holomorphic 
	involution on ${\cal L}_H$. Together with the symmetry properties of the 
	Abelian integrals in the exponent of (\ref{8.16}), 
	this implies that the transformation 
	$\zeta\to-\zeta$ acts as the complex conjugation together with a change 
	of the upper index. Thus we have for the 
	characteristics in (\ref{equ1}) $f(-\zeta)=\bar{f}(\zeta)$.
	\end{proof}
\begin{remark}
If the theta function contains only blocks of the form
	 $\left[\begin{array}{cc}
		0 & 0  \\
		1 & 1
	\end{array}\right]$, the resulting $f$ is just the complex conjugate of 
	the Ernst potential built with the Riemann theta function. This means 
	that the two cases in (\ref{equ1}) are related through  complex 
	conjugation. It is however possible to combine any number of these 
	two blocks in which case the Ernst potential cannot be simply reduced to 
	the case of the Riemann theta function or its complex conjugate. In the 
	case of a rotating body, complex conjugation of the Ernst potential 
	only implies that the angular velocity of the body changes its 
	sign.
\end{remark}

 The above results suggest that it is 
 possible to identify a whole subclass of solutions among (\ref{8.16}) that are 
 asymptotically flat, regular except at a closed contour and equatorially 
 symmetric, i.e.\ solutions that might describe the 
 exterior of a rotating body and might be helpful in the construction of 
 solutions to boundary value problems for the Ernst equation. 
 We get
\begin{theorem}
	Let ${\cal L}_H$ be a regular hyperelliptic surface of even genus $g=2s$ 
	of the form (\ref{3.22a})  without real branch points.  
	Let $\Gamma$ be a closed, 
smooth contour on ${\cal L}_H$ such that 
with $P=(K,\mu(K))\in \Gamma$ also $\bar{P}=(\bar{K},\mu(\bar{K}))\in
\Gamma$ and $(-K,\mu(K))\in \Gamma$ and $E_i\notin\Gamma$. Let there be 
given a finite nonzero function $G$  on 
$\Gamma$ subject to $G(\bar{P})=\bar{G}(P)=G((-K,\mu(K))$. 
Choose the characteristic $\left[\alpha\atop\beta\right]$ such 
that it consists of blocks of the form $\left[
\begin{array}{cc}
	0 & 0  \\
	1 & 1
\end{array}
\right]$ and $\left[
\begin{array}{cc}
	0 & 0  \\
	0 & 0
\end{array}
\right]$ as in theorem 6.2.

Then
\begin{equation}\label{2}
f(\rho,\zeta)=\frac{\Theta\left[\alpha\atop
\beta\right](\omega(\infty^{+})+u)}{\Theta\left[\alpha\atop
\beta\right](\omega(\infty^{+})-u)}
\exp\left\{
\frac{1}{2\pi{\rm i}}\int\limits_\Gamma
\ln G(\tau)\diff\omega_{\infty^{+}\infty^-}(\tau)
\right\}\enspace,
\end{equation}
is\\
1. a regular solution to the Ernst equation for $P_0\notin \Gamma$ if 
condition (\ref{reg}) holds.\\
2. in general discontinuous at 
$\Gamma_z$ given by $P_0\in \Gamma$,\\
3. asymptotically ($|z|\to \infty$) given by $f=1-2m/|z|$ where $m$ is a 
finite real constant if $\Theta'\left[\alpha'\atop\beta'\right](u')\neq0$,\\
4. equatorially symmetric.
\end{theorem}
\begin{proof}
	 From 
	(\ref{8.16}) it can be seen that $f$ is a solution to the Ernst equation.  
	The regularity properties  follow from the previous section. 
	 Asymptotic behaviour and 
	equatorial symmetry follow from above.
\end{proof}

\begin{remark}
	The choice of this class is mainly due to regularity requirements. 
	If all singularities like real branch points or the singularities of 
	the Abelian integrals of the second kind lie within the contour $\Gamma$ 
	where the solution is not considered since the region is 
	assumed to be filled with matter, 
	they would not affect the vacuum region. However this would not enlarge 
	the degrees of freedom (one real--valued function and a set of complex 
	parameters) if one wants to solve boundary value problems.
	\end{remark}
We will discuss the common properties of the solutions in this subclass in 
the following.
\paragraph{Mass and Angular Momentum}
The asymptotic behaviour of the Ernst potential, $f=1-2m/|z|-
2{\rm i}j/|z|^2+\cdots$,
follows already from the axis potential (\ref{sing7}). The equatorial 
symmetry implies that $m$ and $j$ are real constants. In the following it 
is convenient to introduce rescaled coordinates $\tilde{z}=z/R$ where $R$ 
is the radius of the smallest sphere that totally contains the contour 
$\Gamma_z$, and the dimensionless quantities $M=m/R$, $J=j/R^2$. 
This implies that the contour shrinks to a point in the limit 
$R\to0$. If we use the differential operator $D_{\infty^+}$ we get for the 
ADM-mass with (\ref{sing7})
\begin{equation}
	M=\frac{D_{\infty^+}\Theta\left[\alpha'\atop\beta'\right](u')}{
	\Theta\left[\alpha'\atop\beta'\right](u')}+\frac{1}{2p {\rm i}}
	\int_{\Gamma}^{}\ln G \diff\omega'_{1,\infty^+}
	\label{d1}.
\end{equation}
A solution is of course 
only physically acceptable if the mass is positive. 
Similarly one can see that the angular momentum is proportional to $1/
\Theta^2\left[\alpha'\atop\beta'\right](u')$.

\paragraph{Minkowskian limit}

It is possible to parametrize a solution by the mass and the angular 
momentum. For $M<<1$ and $J<<1$, the solution is nearly Minkowskian. 
It can be directly seen that this limit is obtained for $G\to1$. This 
implies that the solutions from above for $|G|\sim 1$ are in the regime of 
small gravitational fields. This is also the regime of the Newtonian limit 
if the solution has one.

\paragraph{Ultrarelativistic limit}

It can be seen from (\ref{d1}) that both the mass and the angular momentum
diverge if $\Theta'\left[\alpha'\atop\beta'\right](u')=0$ but that $M^2/J$
remains finite in this case. This suggests that this divergence is best
understood as the limit $R\to0$ as was already done in \cite{bawa} for 
the case of the rigidly rotating dust disk: in this limit, the 
gravitational fields at the surface $\Gamma_z$ of the matter source become 
so strong that it is hidden behind a horizon, i.e.\ its coordinate radius 
tends to zero. It is suggestive to consider this limit as the 
ultrarelativistic limit of the solution. For finite $R$, the Ernst 
potential is purely imaginary on the axis which means that the solution 
is no longer asymptotically flat. If one takes the limit $R\to0$, one ends 
up with the Kerr solution in this case which gives further support to the
interpretation of this limit as the ultrarelativistic limit (this
interpretation is of course only consistent if the limit gives the extreme Kerr
solution for which the horizon is given in the used coordinates by
$\rho=\zeta=0$). If the limit $R\to0$ is taken for $\rho=\zeta=0$ with
$\rho/\rho_0$ and $\zeta/\rho_0$ finite (this 
corresponds to an observer on the contour that vanishes behind the horizon),
the resulting solution will not be asymptotically flat. 
For an observer on $\Gamma_z$, the exterior region is in infinite geodesic 
distance, and thus completely decouples from the exterior. 
In the context of a boundary value problem this limit corresponds to a stability
limit for the solution: if the boundary data reach a certain critical value,
the solution will no longer be regular.

Thus solutions in (\ref{2}) should be physically interesting in the 
parameter range $0<M/R<\infty$. Notice that the solutions have an analytic 
continuation beyond the upper limit. It can be seen however from (\ref{d1}) 
that the mass changes its sign in this case since 
$\Theta\left[\alpha'\atop\beta'\right](u')$ has zeros of first order. 
Consequently these `overextreme' solutions, that do not have a Newtonian
limit, are probably not physically interesting, at least they are
unacceptable in the region with negative mass.

\section{Reduction of the Ernst potential}

The explicit form of the solutions (\ref{2}) in terms of theta functions 
has a number of advantages in contrast to the linear integral equations to 
which the solution of boundary or initial value problems can be reduced in 
the case of integrable non--linear evolution equations\footnote{There are 
solutions in terms of theta functions for these equations, too. But as we 
have pointed out already, these solutions are always periodic or 
quasiperiodic. The solutions to the Ernst equation discussed here are 
however gauge equivalent to solutions to a linear integral equation as was 
shown in \cite{prd}.}: it is possible to identify physically interesting 
features as ergospheres or the ultrarelativistic limit explicitly. Since 
the theta functions are transcendental, final results normally 
can only be obtained 
numerically. It is however possible to address most features like the 
condition for the occurrence of ergospheres (\ref{ergo}) 
directly without having to determine the Ernst potential numerically in the 
whole spacetime.

The numerical treatment of theta functions is comparatively simple since 
the exponential series converges rapidly due to the factor 
$\exp\left(\frac{1}{2}\pi_{ij}n_in_j\right)$ where $\Re(\pi_{ij})<0$. It is
however obvious that the numerics become more and more tedious the larger 
the genus $g$ of the Riemann surface ${\cal L}_H$ is. Therefore it is an 
important question whether the Riemann surface can be reduced in 
physically interesting cases to surfaces of lower genus. Loosely speaking 
this is possible if there exists a special relation between the branch 
points (see Weierstrass' discussion of the case $g=2$ which is referred to 
in \cite{algebro} 
and references given therein). Since the branch points $P_0$, 
$\bar{P}_0$ are parametrized by the physical coordinates and can thus take 
on arbitrary complex values, such a reduction will only be possible at 
special points of the spacetime which will in general not be of special 
physical interest. A general reduction of the Riemann surface is possible 
if there exist non--trivial automorphisms on the surface. For the class of 
equatorially symmetric solutions discussed here, this is the case in the 
equatorial plane and on the axis. There the surfaces ${\cal L}_H$  and 
$\Sigma'$ have defining equations $\mu(K)$ and $\mu'(K)$ which both depend 
only on $K^2$. Thus on both surfaces there is the involution $T$ defined 
by $(K,\mu(K))\to (-K,\mu(-K))$. 

For the sake of simplicity, we will only discuss the characteristic 
$\alpha_i=\beta_i=0$ and the case $E^1_i=-\bar{E}^2_i$ (the general case 
can be inferred from the resulting relations without problems). 
We will concentrate on disks of radius $\rho_0$ since they 
are an interesting model for galaxies. The Ernst potential simplifies 
in the equatorially 
symmetric case at the disk where the boundary data are prescribed, what
makes disks the most promising objects in the search for solutions to 
boundary value problems to the Ernst equation in closed form. We recall 
that the first solution of such a problem was found for the 
rigidly rotating dust disk \cite{ngbml1}.

In the equatorial plane ($\zeta=0$), the surface ${\cal L}_H$ is then given 
by $\mu^2(K)=(K^2+\rho^2)\prod_{i=1}^{s}(K^2-E_i^2)(K^2-\bar{E}_i^2)$. We cut 
the surface as before which implies $Ta_i^1=a_1^2$, $Tb_i^1=b_i^2$ and 
${\rm d}\omega_i^1(TP)=-{\rm d}\omega_i^2(P)$ with $P\in {\cal L}_H$. The 
Riemann surface $\Sigma_1={\cal L}_H/T$ of genus $s$ 
is then given by 
\begin{equation}
	\mu_1^2(x)=x(x+\rho^2)\prod_{i=1}^{s}(x-E_i^2)(x-\bar{E}_i^2)
        \label{prym1}\enspace.
\end{equation}
The holomorphic differentials $dv_i$ in $\Sigma_1$ dual to $(a_i,b_i)$ (the
projection of the cuts on ${\cal L}_H$ onto $\Sigma_1$) follow from ${\rm d}v_i
={\rm d}\omega_i^1-{\rm d}\omega_i^2$. The so called Prym differentials ${\rm 
d}w_i$ which change the sign under $T$ are given  by ${\rm d}w_i=
{\rm d}\omega_i^1+{\rm d}\omega_i^2$. They are holomorphic differentials 
on the Riemann surface $\Sigma_2$ of genus $s$ with 
\begin{equation}
	\mu_2^2(y)=(y+\rho^2)\prod_{i=1}^{s}(y-E_i^2)(y-\bar{E}_i^2)
        \label{prym2}\enspace,
\end{equation}
which implies that the Prym variety is a Jacobi variety in this case. The 
Riemann matrix on ${\cal L}_H$ has the form 
\begin{equation}
	\Pi=\frac{1}{2}\left(
	\begin{array}{cc}
		\Pi^1+\Pi^2 & \Pi^2-\Pi^1  \\
		\Pi^2-\Pi^1 & \Pi^1+\Pi^2
	\end{array}
        \right)\enspace,
	\label{red3}
\end{equation}
where the $\Pi^i$ are the Riemann matrices on $\Sigma^i$ respectively. The 
theta function on ${\cal L}_H$ thus factorizes into products of theta 
functions on the $\Sigma_i$,
\begin{equation}
	\Theta(x_1|x_2,\Pi)=\sum_{\delta}^{}\Theta\left[\delta\atop0\right]
	(x_1+x_2;2\Pi^2)\Theta\left[\delta\atop0\right]
        (x_1-x_2;2\Pi^1)\enspace,
	\label{red4}
\end{equation}
where each component of the $s$-dimensional vector $\delta$ takes the
values $0,1$. Thus the theta function on the surface of genus $2s$ can be 
expressed via theta functions on surfaces of genus $s$.

In the case of the Ernst potential (\ref{2}), further simplifications 
follow from the fact that $\infty$ is a branch point of $\Sigma_2$. For the 
contour integrals $u$, we obtain for disks
\begin{equation}
        u_v=\frac{1}{\pi {\rm i}}\int_{\Gamma_v}^{}\ln G\diff v,\quad
        u_w=\mbox{sgn}\zeta\frac{1}{\pi {\rm i}}\int_{\Gamma_w}^{}\ln G\diff w
        \enspace,  \label{red11}
\end{equation}
where $\Gamma_v$ is the contour in the $+$-sheet of $\Sigma_1$ between $0$
and $-\rho^2$  along the real axis, and $\Gamma_w$ is the part of the real 
axis in the upper sheet of $\Sigma_2$ between $-\infty$ and $-\rho^2$. The 
formula for $u_w$ shows that it does matter whether the equatorial plane is 
approached from the upper or the lower side (the Ernst potential is not 
regular at the disk). For $\rho>\rho_0$, we have $u_w=0$ ($G=1$ in the 
exterior of the disk). Similarly we get for the integral in the exponent
of (\ref{2})
\begin{equation}
        I_v\doteq\frac{1}{2\pi {\rm i}}\int_{\Gamma}^{}\ln G\diff
        \omega_{\infty^+\infty^-}=
        \frac{1}{2\pi {\rm i}}\int_{\Gamma_v}^{}\ln G\diff v_{\infty^+\infty^-}
        \label{red12}\enspace.
\end{equation}
Summing up we can write the Ernst potential in the equatorial plane in the 
form 
\begin{equation}
	f=\frac{\sum_{\delta}^{}\Theta\left[\delta\atop0\right]
	(v(\infty^+)+u_v;2\Pi^1)\Theta\left[\delta\atop\beta\right]
	(u_w;2\Pi^2)}{\sum_{\delta}^{}\Theta\left[\delta\atop0\right]
	(v(\infty^-)+u_v;2\Pi^1)\Theta\left[\delta\atop\beta\right]
        (u_w;2\Pi^2)}{\rm e}^{I_v}\enspace,
	\label{prym3}
\end{equation}
where $\beta_i=1$, and where $v(P)=\int_{-\rho^2}^{P}dv$. 
The reality properties of the above theta functions 
imply together with (\ref{red11}) the condition for equatorial symmetry 
$f(-\zeta)=\bar{f}(\zeta)$. Thus the imaginary part of $f$ jumps at the 
disk. For $\rho>\rho_0$ (where $u_w=0$), only the terms with even  
characteristics in 
(\ref{prym3}) will survive which leads to a real Ernst potential. This 
implies that the Ernst potential is regular outside the disk as it should 
be. The formula (\ref{prym3}) can also be used to determine asymptotic 
quantities as angular momentum and ADM-mass in the limit of
$\rho\to\infty$ as was done previously on the axis.

A similar reduction as in the equatorial plane is possible on the axis. 
There the Riemann surface $\Sigma'$ also has the involution $T$ which 
makes it possible to factorize the surface into the surfaces $\Sigma_1'$ 
and $\Sigma_2$ where the $\Sigma_i$ are as above and where $\Sigma_1'$ 
is $\Sigma_1$  with the cut 
$\left[0,-\rho^2\right]$ removed. Thus the theta function $\Theta'$ 
on the surface  $\Sigma'$ of 
genus $2s-1$ can be expressed via theta functions on surfaces of genus 
$s-1$ and $s$ respectively. In the case $g=2$, this will not lower the 
genus of the Riemann surfaces under consideration on the axis.
We will not give the Ernst potential on the axis
since it will not be used here (the formula is helpful if one wants to 
calculate the multipole moments on the axis for higher genus).

\section{The case $g=2$}

The simplest non-static solutions within the class (\ref{2}) are of genus
2 since solutions of genus 0 belong to the Weyl-class. Interestingly the
first solution to a physically relevant boundary value problem, 
the rigidly rotating dust disk \cite{ngbml1} with dust parameter $\nu=
2\Omega^2\rho_0^2{\rm e}^{-2V_0}$ where $\Omega$ is the angular velocity 
in the disk, and  ${\rm e}^{-V_0}-1$ is the central redshift and radius 
$\rho_0$,  belongs to
this subclass. There the characteristic is $\left[
\begin{array}{cc}
	0 & 0  \\
	1 & 1
\end{array}
\right]$, the branch points are given by 
$E=\sqrt{{\rm i}/\nu-1}$, and the function $G$ has the form $G=\left(
\sqrt{1+\nu^2(K^2+1)^2}+\nu(K^2+1)\right)^2$ where we have used 
dimensionless coordinates $\rho/\rho_0$ and $\zeta/\rho_0$.  
Therefore we will discuss the case $g=2$ as an example in more detail.

Solutions (\ref{2}) of genus 2 will be regular except at the contour 
$\Gamma_z$ if $\Theta(\omega(\infty^-)+u)\neq 0$.  This is 
equivalent to the condition that the divisor $A$, defined by the Jacobi 
inversion problem $\omega(A)-\omega(D)=u$, does not contain both $\infty^-$ 
and $X$ where $X=P_0$ or $X=\bar{P}_0$. This implies that the equations
\begin{eqnarray}
	\int_{-E}^{\infty^-}\frac{{\rm d}\tau}{\mu(\tau)}
	+ \int_{\bar{E}}^{X}\frac{{\rm d}\tau}{\mu(\tau)}& = & 
	\frac{1}{2\pi {\rm i}}\int_{\Gamma}^{}\frac{\ln G{\rm d}\tau}{\mu(\tau)}
	\label{g2.1} \\
	\int_{-E}^{\infty^-}\frac{\tau{\rm d}\tau}{\mu(\tau)}
	+ \int_{\bar{E}}^{X}\frac{\tau{\rm d}\tau}{\mu(\tau)}& = & 
	\frac{1}{2\pi {\rm i}}\int_{\Gamma}^{}\frac{\ln G\tau{\rm d}\tau}{\mu(\tau)}
	\label{g2.2}
\end{eqnarray}
must not hold simultaneously for all $\rho$ and $\zeta$ in the vacuum 
region. Since the limits of the integration are fixed, this inversion 
problem will in general not have a solution. The equations constitute a
relation between the physical parameters that characterize the solution. In 
the case of the dust disk, this is the parameter $\nu$ and the radius 
$\rho_0$. This implies that 
the above condition will determine the allowed parameter range for $\nu$ 
for which there are no further singularities in the whole spacetime except the 
disk. 

The condition for ergospheres can be obtained from a similar system of 
equations as above: simply replace $\infty^-$ in (\ref{g2.1}) and 
(\ref{g2.2}) by a point on ${\cal L}_H$ which must not be 
$\infty^-$ or $X$ but is otherwise arbitrary. Then an ergosphere occurs if both 
equations hold simultaneously which gives in the above example the values 
for $\nu$ for which there exists a non--empty set of points $\rho$ and 
$\zeta$, the ergosphere. Since one of the points in the divisor $A$ is in 
this case essentially arbitrary, the conditions for ergospheres will be 
satisfied much more frequently than the condition for a singularity where 
both points of $A$ are prescribed.

The ergosphere has only common points with the axis 
in the ultrarelativistic limit which is 
given by $\vartheta_4(u_1')=0$ (we use the notation for elliptic theta 
functions of \cite{erdelyi}). This condition is equivalent to
\begin{equation}
	\frac{1}{2\pi {\rm i}}\int_{\Gamma}^{}\frac{\ln G {\rm d}\tau}{\mu'(\tau)}
	=(2n+1)\int_{-E}^{E}\frac{{\rm d}\tau}{\mu'(\tau)}
	\label{g2.3}
\end{equation}
where $n=0,1,2,\ldots$. Since this relation is independent of the physical 
coordinates, it determines the values for $\nu$ at which the 
ADM-mass diverges. Together with the conditions (\ref{g2.1}) and
(\ref{g2.2}), this determines the allowed parameter range for $\nu$: the 
absence of singularities except the disk and the fact that the mass shall 
vary between 0 and $\infty$ (the ultrarelativistic limit). 

In the case $g=2$, the potential in the equatorial plane and on the axis 
can be expressed in terms of elliptic theta functions. The formula for the 
axis reads 
\begin{eqnarray}
	f(\rho=0,\zeta)&=&\frac{\vartheta_4(\omega_1'|_{\zeta^+}^{\infty^+}+u_1')
	\vartheta_1(\omega_1'|_{\zeta^-}^{\infty^+})+
	\exp(-u_2)\vartheta_4(\omega_1'|_{\zeta^-}^{\infty^+}+u_1')
	\vartheta_1(\omega_1'|_{\zeta^+}^{\infty^+})}{
	\vartheta_4(\omega_1'|_{\zeta^+}^{\infty^+}-u_1')
	\vartheta_1(\omega_1'|_{\zeta^-}^{\infty^+})+
	\exp(u_2)\vartheta_4(\omega_1'|_{\zeta^-}^{\infty^+}-u_1')
	\vartheta_1(\omega_1'|_{\zeta^+}^{\infty^+})}
	\nonumber\\
	&&\exp\left\{\frac{1}{2\pi {\rm i}}
        \int\limits_\Gamma
\ln G(\tau)\diff\omega'_{\infty^{+}\infty^-}(\tau)+u_2\right\}
	\label{7}.
\end{eqnarray}
In the equatorial plane we have
\begin{equation}
	\bar{f}=\frac{\vartheta_3(v+u_v;2\Pi^1)\vartheta_4(u_w;2\Pi^2)
	-\vartheta_2(v+u_v;2\Pi^1)\vartheta_1(u_w;2\Pi^2)}{
	\vartheta_3(u_v-v;2\Pi^1)\vartheta_4(u_w;2\Pi^2)
	+\vartheta_2(u_v-v;2\Pi^1)\vartheta_1(u_w;2\Pi^2)}e^{I_v}
	\label{red13}
\end{equation}
where $v=\int_{-\rho^2}^{\infty^+}\diff v$. For $\rho>\rho_0$, the exterior of
the disk, we get
\begin{equation}
	f=\bar{f}=\frac{\vartheta_3(v+u_v;2\Pi^1)}{\vartheta_3(
	u_v-v;2\Pi^1)}e^{I_v}.
	\label{red14}
\end{equation}
In both cases the formulae for a different characteristic are obtained by 
complex conjugation.

The above relations illustrate that important features of the solutions of 
genus 2 can be discussed with the help of the standard elliptic theory. It 
is thus an interesting question which boundary value problems lead to 
solutions within this subclass.

\section{Outlook}

In this paper, it was shown that it is possible to identify within a class
of hyperelliptic solutions to the Ernst equation a subclass whose solutions 
could describe the exterior of a body of revolution: they are asymptotically 
flat, equatorially symmetric and regular except at the surface of the 
body. This subclass could consequently be interesting in the context of 
boundary value problems for the Ernst equation. There the boundary data are 
either induced by an interior solution or a surfacelike matter distribution 
as in the case of the dust disk: in the general case, one would have to 
fix two real functions at the boundary in order to satisfy the boundary 
conditions. Within the class considered here, 
one has the freedom to choose one real valued function 
($G$) and a set of free parameters $E_i$, the branch points of the Riemann 
surface. Whether the subclass discussed here can actually be used to  
solve boundary value problems, and when these degrees of freedom will be 
indeed sufficient is an open question. However it is remarkable 
that it is possible to identify a whole generic subclass of regular 
equatorially symmetric solutions. This gives reasonable hope that further 
solutions to physically interesting boundary value problems may be found 
within the class of hyperelliptic solutions to the Ernst equation.\\[1ex]
{\bf Acknowledgement:}\\
We thank H.~Farkas, J.~Frauendiener and H.~Pfister for helpful discussions 
and hints. One of us (CK) acknowledges support by the DFG.

\end{document}